\documentclass{elsarticle}
\usepackage{amsmath,amssymb,amsthm,amsfonts}
\usepackage{amssymb}
\newtheorem{thm}{Theorem}[section]

\newtheorem{lem}{Lemma}[section]

\newtheorem{rem}{Remark}[section]
\theoremstyle{definition}

\numberwithin{equation}{section}
\DeclareMathSymbol{\C}{\mathalpha}{AMSb}{"43}

\newcommand{\eps}{\varepsilon}

\newcommand{\lam}{\lambda}

\newcommand{\R}{{\mathbb{R}}}
\newcommand{\h}{{\mathcal{H}}}
\newcommand{\inte}{\int_{\mathbb{R}^2}}

\newcommand{\bsub}{\begin{subequations}}
\newcommand{\esub}{\end{subequations}$\!$}

\begin{document}

\begin{frontmatter}

\title{Blow-up behavior of ground states for a  nonlinear Schr\"{o}dinger system  with attractive and  repulsive  interactions
}


 \author[WIPM]{Yujin Guo}
\ead{yjguo@wipm.ac.cn}
\author[WUT]{Xiaoyu Zeng}
\ead{xyzeng@whut.edu.cn}
   \author[WUT]{Huan-Song Zhou \corref{cor}}
   \ead{hszhou@whut.edu.cn}
   \cortext[cor]{Corresponding author.} 
   \cortext[cor]{to appear in J. Differential Equations (2017), https://doi.org/10.1016/j.jde.2017.09.039 (article in press)}
   \address[WIPM]{Wuhan Institute of Physics and Mathematics,  Chinese Academy of Sciences \\P.O. Box 71010, Wuhan 430071, P.R. China}
   \address[WUT]{Department of Mathematics, Wuhan University of Technology, Wuhan 430070, China}


\date{}
\begin{abstract}
We consider a nonlinear Schr\"odinger system arising in a two-component Bose-Einstein condensate (BEC)  with attractive intraspecies interactions and repulsive interspecies interactions in $\R^2$.  We get ground states of this system by solving  a constrained minimization problem. For some kinds of trapping potentials, we prove that the minimization problem has a minimizer if and only if the attractive  interaction strength $a_i (i=1,2)$ of each component of the BEC system is strictly less than a threshold $a^*$.
Furthermore,  as $(a_1, a_2)\nearrow (a^*, a^*)$,
the asymptotical behavior  for the minimizers of the  minimization problem is discussed. Our results show that each component of the BEC system concentrates at a global minimum of the associated trapping potential.
\end{abstract}
\end{frontmatter}


\section{Introduction}

We are concerned with the  ground states of the following nonlinear Schr\"{o}dinger system:
\begin{equation}\label{sys}
\begin{cases}
 -\Delta u_{1}+V_1(x)u_{1}=\mu_{1} u_{1}+a_1
u_{1}^3-\beta u_{2}^2u_{1}\quad \mbox{in}\,\ \R^2,\\
-\Delta u_{2}+V_2(x)u_{2}=\mu_{2} u_{2}+a_2 u_{2}^3-\beta
u_{1}^2u_{2}\quad \mbox{in}\,\ \R^2.
\end{cases}
\end{equation}
System (\ref{sys}) is an important model for a two-component Bose-Einstein condensate (BEC), see e.g., \cite{Bao,HME,MMR,MBG}, where
$(V_1(x),V_2(x))$ is the external potential, $(\mu_1,\mu_2)\in\R \times \R$ is the chemical potential, $a_i>0$ ($i=1,\,2$) represents that the intraspecies interactions among  cold atoms are attractive in each component, and $\beta$ denotes the interspecies interaction strength between two components of
the system.  The case of $\beta <0$ (i.e., the attractive interspecies) is recently discussed in \cite{GZZ2}. As a continuation of \cite{GZZ2}, in this paper we focus on the case of the repulsive interspecies interaction, that is, $\beta>0$.

The intraspecies interactions among the atoms in each component of the BEC system can be attractive ($a_i > 0$) or repulsive ($a_i < 0$). The case of repulsive intraspecies was analyzed recently in \cite{Bao,KL,MA,R} and the  references therein. Since the
 intraspecies interaction is attractive in our case (i.e. $a_i>0$), we may expect that each BEC component  collapses if the particle number increases beyond a critical value by the inspiration of  the single component BEC, see \cite{D,GS,GZZ,GZZ2}, etc. Note that there are  interspecies interactions
between two components, then a two-component BEC should present more complicated phenomena  than a single-component BEC, and it is more difficult to discuss in detail. Our main interest in the present paper is to investigate a two-component BEC with  attractive intraspecies interactions and repulsive interspecies interactions. In our previous  work \cite{GZZ2}, we analyze a two-component BEC with both attractive interactions of intraspecies and interspecies. For problem (\ref{sys}) with $V_i(x) (i=1,2)$ being constants, we refer the reader to the paper \cite{LW0} and the references therein.


 It is known that (\ref{sys}) is also  the Euler-Lagrange equations for the following constrained minimization problem
\begin{equation}\label{eq1.3}
e(a_1,a_2):=\inf_{\{(u_1,u_2)\in \mathcal{X},\int_{\R^2}
|u_1|^2dx=\int_{\R^2}
|u_2|^2dx=1\}} E(u_1,u_2),
\end{equation}
where  $ E(u_1,u_2)$  is defined on $\mathcal{X}$ by
\begin{equation}\label{f}
  E(u_1,u_2):=\sum_{i=1}^2\int_{\R ^2} \Big(|\nabla
  u_i|^2+V_i(x)|u_i|^2-\frac{a_i}{2}|u_i|^4\Big)dx+\beta\inte u_1^2u_2^2dx .
\end{equation}
 The space $\mathcal{X}=\h_1\times \h_2$ and $\h_i = \Big \{u\in  H^1(\R ^2):\ \int _{\R ^2}  V_i(x)|u(x)|^2 dx<\infty \Big\}$ with
\begin{equation*}
\begin{split}
   \|u\|_{_{\h_i}}=\Big(\int _{\R ^2} \Big[|\nabla u|^2+ V_i(x)|u(x)|^2\Big] dx\Big)^{\frac{1}{2}}, \,\ i=1,\,2.
   \end{split}
\end{equation*}
Throughout the paper, we let  $a_i>0$ ($i=1,\,2$), $\beta >0$ and let $V_i(x)$  satisfy
\begin{equation}\label{1.12}
V_i(x)\in L^\infty_{\rm
loc}(\R^2),\,\ \lim_{|x|\to\infty} V_i(x) = \infty\,\ \text{and}\,\ \inf_{x\in
\R^2} V_i(x) =0,\,\ i=1,\,2.
\end{equation}
By a  similar analysis to that of \cite{GWZZ}, where  a  single-component minimization problem was considered, we see that if $(u_1,u_2)$ is a minimizer of (\ref{eq1.3}) and $(a_1,a_2,\beta)$ locates in a   suitable region, then $(u_1,u_2)$ is also a ground state of (\ref{sys}) for some Lagrange multiplier  $(\mu_1,\mu_2)\in \R\times\R$, which is usually called as a pair of eigenvalues. In this paper, we only consider the ground states of (\ref{sys}) obtained by the minimizers $(u_1,u_2)$ of the minimization problem (\ref{eq1.3}), that is, $(u_1,u_2)$ satisfies \eqref{sys} with eigenvalues  $\mu_i\in \R \ (i=1,2)$ because of the $L^2$-norm constraints. Generally, $\mu_i$ are not removable by rescaling since the presence of $V_i(x)$. However, by using \eqref{eq3.28}, \eqref{eq3.30} and Theorem 3.1  in Section 3, we may have some detailed estimates on the eigenvalues of $\mu_i$.


For discussing the existence and nonexistence of minimizers for $e(a_1,a_2)$, we need to recall some results for the following nonlinear scalar field
equation
\begin{equation}
-\Delta u+ u-u^3=0\  \mbox{  in } \  \R^2,\  \mbox{ where }\  u\in
H^1(\R ^2).  \label{Kwong}
\end{equation}
It is well-known from \cite{GNN,K,Li,mcleod} that (\ref{Kwong}) admits a unique positive solution (up to
translations), which
can be taken to be radially symmetric about the origin and we
denote it  by $Q=Q(|x|)$. Moreover, we denote
\begin{equation} \label{a*}
a^*:=\|Q\|_2^2=\inte |Q(x)|^2dx,
\end{equation}
and, in what follows, we always denote by $\|\cdot\|_p$ with $p\in (1,+\infty)$ the norm of $L^p(\mathbb{R}^2)$.
 Now, we may recall from \cite{W} the
following Gagliardo-Nirenberg inequality
\begin{equation}\label{GNineq}
\inte |u(x)|^4 dx\le \frac 2 {a^*} \inte |\nabla u(x) |^2dx
\inte |u(x)|^2dx ,\   \  u \in H^1(\R ^2),
\end{equation}
and ``$=$" holds if  $u(x) = Q(|x|)$. Moreover, it follows  from
(\ref{Kwong}) and (\ref{GNineq}) that 
\begin{equation}\label{1:id}
\inte |\nabla Q |^2dx  =\inte Q ^2dx=\frac{1}{2}\inte Q ^4dx ,
\end{equation} 
and by \cite[Prop.~4.1]{GNN} we also know that
 \begin{equation}
Q(x) \, , \ |\nabla Q(x)| = O(|x|^{-\frac{1}{2}}e^{-|x|}) \quad
\text{as  $|x|\to \infty$.}  \label{4:exp}
\end{equation}
The inequality (\ref{GNineq}) is used in \cite{Bao,GS} to study 
the following minimization problem
\begin{equation}\label{1:two}
e_i(a):=\inf_{\{u\in\h_i,\int_{\mathbb{R}^2} u^2dx=1\}}
E_{a}^i(u),
\end{equation}
where
\begin{equation}
E_{a}^i(u):=\int_{\R ^2} \big(|\nabla
  u(x)|^2+V_i(x)|u(x)|^2\big)dx-\frac{a}{2}\int_{\R ^2}|u(x)|^4dx, \  u\in \h_i,\ i=1,\,2.\label{single}
\end{equation}
Using  the inequality (\ref{GNineq}) and the concentration procedures, it was proved in \cite{Bao,GS} that (\ref{1:two}) can be attained if and only if $a<a^*:=\|Q\|_2^2$.
%
Motivated by \cite{Bao,GS}, we have the following theorem on the  existence and non-existence  of minimizers for problem (\ref{eq1.3}).

\begin{thm}\label{thm1}
{\em Let $Q$ and $a^*$ be given by (\ref{a*}). Suppose  $\beta>0$ and $V_i(x)$ ($i=1,2$) satisfies  (\ref{1.12}). 
Then
\begin{enumerate}
\item [ (i)] If $0\leq a_i< a^*$ ($i=1$ and $2$), there exists at least one minimizer for (\ref{eq1.3}).
\item [(ii)]If  either $a_i > a^*$ ( $i=1 ~or~ 2)$, or $a_1=a_2=a^*$, there does not exist any minimizer for (\ref{eq1.3}).
\end{enumerate}
Moreover, $e(a_1, a_2) > 0$ for  $0\leq a_1,a_2<a^*$,
$\lim_{(a_1,a_2)\nearrow (a^*,a^*)} e(a_1, a_2) = e(a^*, a^*) = 0$, and
$e(a_1,a_2) = -\infty$ for either $a_1>a^*$ or $a_2>a^*$. }
\end{thm}

After we finish the first draft  of the paper, we note that the similar results on the existence of minimizers already appeared  in \cite{Bao}. We mention that the proof of Theorem \ref{thm1} (i) is similar to the single component BEC problem \cite{Bao,GS}.  However, for part (ii), especially when $a_1=a_2=a^*$, because  of the presence  of the  term $-\beta\inte|u_1|^2|u_2|^2dx$ in $E(\cdot, \cdot)$ the arguments used in \cite{Bao,GS,GZZ2}  are not able to  give the optimal energy estimate for (\ref{eq1.3}), thus we cannot directly use their methods  to obtain  the non-existence of minimizers. To overcome this difficulty, we need to modify the trial function used in \cite{GS} and  make some refined calculations for the energy of (\ref{eq1.3}).

Without loss of generality, in what  follows we may restrict the minimization of (\ref{eq1.3}) to non-negative vector functions $(u_1,u_2)$, since $E(u_1,u_2)\geq
E(|u_1|,|u_2|)$ holds for any $(u_1,u_2)\in \mathcal{X}$, due to the fact that $\nabla |u_i|\leq |\nabla u_i|$ a.e. in $\R^2$, where $i=1,\,2$.

Inspired by the work of \cite{GZZ2}, our main results of the  paper are focused on  the limit behavior of nonnegative minimizers $(u_{a_1},u_{a_2})$ of
(\ref{eq1.3}) as $(a_1,a_2)\nearrow (a^*,a^*)$ in  the case of $\beta>0$. Since now the intraspecies and interspecies interactions in the BEC system  are not of the same signs, this causes   that the asymptotic  analysis in the present paper  is much more complicated  than  that of \cite{GZZ2} where  only the attractive  interactions (i.e., $a_1,a_2>0$ and $\beta<0$) are involved.
Since
$e(a^*,a^*)=0$, the proof of Theorem 1.1 yields that $\int_{\R^2} V_i(x) |u_{a_i}(x)|^2 dx \to
0 = \inf_{x\in \R^2} V_i(x)$ as $(a_1,a_2)\nearrow (a^*,a^*)$, and hence the  behavior of minimizers  depends on the local profile of $V_i$ near its minima. For this reason,  we simply consider $V_i(x)$ to be of the form
\begin{equation}\label{1:V}
V_i(x)=|x-x_i|^{p_i},\ \   p_i>0, \end{equation} and set
\begin{equation}\label{1:var}
\varepsilon_i:=(a^*-a_i)^\frac{1}{p_i+2},\ i=1,\,2.
\end{equation}
 We first recall the   following results on problem (\ref{1:two})  given in \cite{GS,GZZ}, and its proof is sketched in the appendix for the reader's convenience.

\vskip 0.1truein

\noindent{\bf Proposition A.}   {\em Let $V_i(x)$  be given by
(\ref{1:V}),  then for problem (\ref{1:two}) we have
\begin{enumerate}
\item [\rm(i)]
\begin{eqnarray}\label{1:e}
e_i(a_i)\sim
(a^*-a_i)^\frac{p_i}{p_i+2}=\varepsilon_i^{p_i}\ \ {\rm as}\,\
a_i\nearrow a^*, \ i=1,\,2,
\end{eqnarray}
where $f\sim g$ means that $f/g$ is
bounded from below and above.

\item [\rm(ii)] Let $\bar u_i(x)>0$ ($i=1,2$) be a
positive   minimizer of (\ref{1:two}), then $\bar
u_i(x)$ has only one local (and hence global) maximum point which
approaches $x_i$ as $a_i\nearrow a^* $. Moreover, for any $R>0$,
there exists $C(R)>0$ such that
\begin{equation}\label{1:dec} \bar u_i(x)\leq C(R)
e^{\frac{-\delta|x-x_i|}{\varepsilon_i}} \ \ \text{in}\,\
\R^2\setminus B_R(x_i),
\end{equation}
where $\delta>0$ is independent of $x_i$, $R$ and $\varepsilon_i$.
\end{enumerate}}

Without loss of generality, we assume that $\varepsilon_1\geq\varepsilon_2> 0$ (defined in (\ref{1:var})), let \begin{equation}\label{1:del}
 \delta_0:=\delta
|x_1-x_2|,\ \ \text{where}\,\   \delta >0\,\  \text{is given by}\,\
(\ref{1:dec}),\end{equation}
and consider the upper limit
\begin{equation}\label{1:finity}
0\leq  \mathbf{L}:=\overline{\lim}_{(a_1, a_2)\nearrow (a^*,a^*)}\frac{e^{ -\delta_0 (a^*-a_1)^{-\frac{1}{p_1+2}} }}{(a^*-a_2)^\frac{p_2}{p_2+2}}\le \infty .
\end{equation}
When the minima of $V_1$ and $V_2$ are not the same, we then have

\begin{thm}\label{thm2}
Let $\beta>0$ and $V_i(x)$ be given by (\ref{1:V}) and
$x_1\neq x_2$. Suppose $(u_{a_1},u_{a_2})$ is a
non-negative minimizer of (\ref{eq1.3}), and assume $\eps_1\geq\eps_2>0$ with $\eps_i$ given by (\ref{1:var}). Then each $u_{a_i}$ has a unique maximum point $x_{a_i}$ which approaches to $x_i$ as
$(a_1,a_2)\nearrow (a^*,a^*)$. Moreover,
\begin{itemize}
\item[\rm(i).] $u_{a_1}$ has the following property:
 \begin{equation*}
\lim_{(a_1, a_2)\nearrow (a^*,a^*)} (a^*-a_1)^\frac{1}{p_1+2} u_{a_1}\Big((a^*-a_1)^\frac{1}{p_1+2}
x+x_{a_1}\Big)=\frac{\lambda_1}{\|Q\|_{2}}Q(\lambda_1x)
 \end{equation*}
 strongly in $H^1(\R^2)$, where $x_{a_1}$ satisfies
 \begin{equation*}
\frac{|x_{a_1}-x_1|}{(a^*-a_1)^\frac{1}{p_1+2}}\to 0 \ \ \text{as}\,\
(a_1, a_2)\nearrow (a^*,a^*),
\end{equation*}
 and $\lam _1>0$ satisfies
\begin{equation*}
  \lambda_1 = \Big(
\frac{p_1}{2}\inte | x|^{p_1}Q^2(x)dx\Big)^\frac{1}{p_1+2}.
\end{equation*}

\item[\rm(ii).] If    (\ref{1:finity}) holds for $0\le \mathbf{L}<\infty$, then
$u_{a_2}$ satisfies
\begin{equation}\label{1:azero}
\lim_{(a_1, a_2)\nearrow (a^*,a^*)} (a^*-a_2)^\frac{1}{p_2+2} u_{a_2}\Big((a^*-a_2)^\frac{1}{p_2+2}
x+x_{a_2}\Big)=\frac{\lambda_2}{\|Q\|_{2}}Q(\lambda_2|x|)
 \end{equation}
 strongly in $H^1(\R^2)$ for some $\lam
 _2>0$. Specially, if  $\mathbf{L}=0$,
(\ref{1:azero}) then  holds for
\[
 \frac{|x_{a_2}-x_2|}{(a^*-a_2)^\frac{1}{p_2+2}}\to 0 \ \ \text{as}\,\
(a_1, a_2)\nearrow (a^*,a^*),\]
and
\[\lambda_2 = \Big( \frac{p_2}{2}\inte |
x|^{p_2}Q^2(x)dx\Big)^\frac{1}{p_2+2}.
\]

\item[\rm(iii).]  If    (\ref{1:finity}) holds for $  \mathbf{L}=\infty$, then
$u_{a_2}$ satisfies
\begin{equation*}
\lim_{(a_1,a_2)\nearrow (a^*,a^*)} \tilde\varepsilon_2
u_{a_2}(\tilde\varepsilon_2
x+x_{a_2})=\frac{\lambda_2}{\|Q\|_{2}}Q(\lambda_2x)
 \end{equation*}
 strongly in $H^1(\R^2)$ for some $\lam _2>0
$, where $\tilde\eps_2>0$ is given and satisfies
\begin{equation*} \tilde\eps_2:=
\Big(\int_{\R^2}u_{a_2}^4dx\Big)^{-\frac{1}{2}}\to 0
\ \ \text{as}\,\ (a_1,a_2)\nearrow (a^*,a^*).
\end{equation*}
\end{itemize}
\end{thm}

When the minima of $V_1$ and $V_2$ are not equal, Theorem \ref{thm2} gives the refined estimate for the limit behavior of
$u_{a_1}(x)$ as $(a_1,a_2)\nearrow (a^*,a^*)$, which is the same as that of the single-component problem discussed in \cite{GS}. On the other hand,  the refined estimates for the limit behavior of
$u_{a_2}(x)$ can be obtained only if $\mathbf{L}=0$. We note that such an additional assumption is true for the case where either $a^*-a_1\sim a^*-a_2$ or
$\eps_2\sim \eps_1^s $ (for some $0<s<\infty$) as $(a_1,a_2)\nearrow (a^*,a^*)$. We also mention  that the proof of Theorem \ref{thm2} is mainly motivated by \cite{GZZ2} where the case $\beta<0$ is considered. In \cite{GZZ2}  we can prove that the two components   must blow up at the same rate if all the interactions in the  system are  attractive. Thus,  by a suitable scaling, the  Euler-Lagrange equation that the minimizers satisfies  can be transformed to the form like (\ref{1:sysytem}) where only one parameter $\eps>0$ exists, and then the arguments in \cite{LW1,LW2,MPS}  can be  partly employed to investigate the limit behavior of minimizers.  However, in our Theorem \ref{thm2}, $u_{a_1}$ and $u_{a_2}$ in general  blow up in different rates,  after the rescaling, there contains two small parameters $\eps_1$ and $\eps_2$ in the  Euler-Lagrange equation, see e.g., (\ref{3:222}) below. Thus, the blow-up analysis becomes more complicated because of the involvement of   $\eps_1$ and $\eps_2$, which go to zero   in different orders.

When the minima of $V_1$ and $V_2$ are the same, the limit behavior  of non-negative minimizers $(u_{a_1},u_{a_2})$ as $(a_1, a_2)\nearrow (a^*,a^*)$ seems  more
complicated. Actually, in this case we note from Theorem \ref{thm1} that both $u_{a_1}$ and $u_{a_2}$ prefer to concentrate at the common minimum point of $V_1$ and $V_2$. However, it is expected that  both $u_{a_1}$ and $u_{a_2}$ repel each other to reduce the energy of the system, due to the fact that the interspecies interaction between the components is repulsive.    For the special case where
\begin{equation}\label{eq1.18}
V_1(x)=V_2(x)=|x-x_0|^p \ \text{ with }p>0,
\end{equation}  the following theorem shows that the concentration rate is reduced essentially  by the effect of the repulsive interspecies interaction.

\begin{thm}\label{thm3}
Suppose $\beta >0$ and (\ref{eq1.18}) holds. Let $(u_{a_1},u_{a_2})$ be a non-negative minimizer of (\ref{eq1.3}) with $(a^*-a_1)\thicksim (a^*-a_2)$
as $(a_1,a_2)\nearrow (a^*,a^*)$. Then, each $u_{a_i}$ has a unique maximum point $x_{a_i}$ which approaches  $x_0$ as $(a_1,a_2)\nearrow (a^*,a^*)$. Moreover, we have for $i=1,\,2,$
\begin{equation*}
\lim_{(a_1,a_2)\nearrow (a^*,a^*)} \epsilon_i(a_1,a_2) u_{a_i}\big(\epsilon_i(a_1,a_2)
x+x_{a_i}\big)=\frac{\lambda_i}{\|Q\|_{2}}Q(\lambda_ix)
\end{equation*}
strongly in $H^1(\R^2)$ for some $\lam _i>0$, where
\begin{equation}\label{eq1.20}
\epsilon_i(a_1,a_2):=
\Big(\int_{\R^2}u_{a_i}^4dx\Big)^{-\frac{1}{2}}\to 0
\ \ \text{as}\,\ (a_1,a_2)\nearrow (a^*,a^*),
\end{equation}
and $x_{a_1}$ and $x_{a_2}$ satisfy  \begin{equation}\label{thm3:3}
\lim_{(a_1,a_2)\nearrow (a^*,a^*)}\frac{|x_{a_1}-x_{a_2}|}{\epsilon_i(a_1,a_2)}=+\infty,  \ \
\limsup_{(a_1,a_2)\nearrow (a^*,a^*)}\frac{|x_{a_i}-x_0|}{\epsilon_i(a_1,a_2)|\ln
\epsilon_i(a_1,a_2)|}<+\infty.
\end{equation}
\end{thm}

\begin{rem}
Different from Theorem \ref{thm2}, in Theorem \ref{thm3} we can only prove that $u_{a_i}$ ($i=1,2$) must blow up as $(a_1,a_2)\nearrow (a^*,a^*)$, but we cannot give the accurate  blow up rate in (\ref{eq1.20}).
\end{rem}

When $V_1(x)$ and $V_2(x)$ share  multiple global minimum points $\{x_1,\cdots ,x_n\}$ with $n\ge 2$, the limit behavior of minimizers as $(a_1,a_2)\nearrow (a^*,a^*)$ is more dramatic. For example, if $V_1(x)=V_2(x)=\prod_{i=1}^n |x-x_i|^{p}$ ($p>0$) has the symmetry, where $x_1,\cdots,x_n$ are arranged on the vertices
of a regular polygon, stimulated by \cite{GS} and Theorem \ref{thm3}, we may expect that the symmetry breaking occurs as $(a_1,a_2)\nearrow (a^*,a^*)$, and the concentration points of both components are unique and however different. As another example, if $V_1(x)=V_2(x)=\prod_{i=1}^n |x-x_i|^{p_i}$, where $p_i>0$ and $p_i\not =p_j$ for $i\not =j$, we may expect from Theorem \ref{thm3} and \cite{GS} that both components concentrate  at the same point $x_0$, which  is the flattest global minimum point of the trap potential, i.e., $x_0$ satisfies
$$x_0=\big\{x_i:\, p_i\ge p_j\ \, \mbox{for\ \, all} \ \, j=1,\cdots n\big\}.$$
We leave the analysis of above different situations to the interested readers.

The concentration phenomena in multiple-component BECs were also analyzed elsewhere in different contexts. For example, positive ground
state solutions of the elliptic system
\begin{equation}\label{1:sysytem}
\begin{cases}
 -\eps^2\Delta
 u_1+V_1(x)u_1=\mu_1u_1^3+\beta u_2^2u_1\quad \mbox{in}\quad \Omega\subset \R^N,\\
 -\eps^2\Delta u_2+V_2(x)u_2=\mu_2u_2^3+\beta u_1^2
 u_2\quad \mbox{in}\quad \Omega\subset \R^N,
\end{cases}
\end{equation}
were discussed as $\varepsilon\to 0^+$, where $\Omega$ is a bounded (or unbounded) domain in $\R^N$, see \cite{LW1,LW2,MPS} and the  references therein.  

This paper is organized as follows: in
Section 2  we prove Theorem \ref{thm1} on the existence and nonexistence of
minimizers for (\ref{eq1.3}). In Section 3 we first establish some energy estimates, upon  which the proof of Theorem \ref{thm2} is addressed.
The proof of Theorem \ref{thm3} is then given in Section 4. In the appendix, the above Proposition A as well as Lemma A  required in showing  Theorem \ref{thm3} are proved.

\section{Existence of Minimizers}
This section is focussed on the proof of Theorem \ref{thm1}, and our proof is inspired by  \cite{Bao,GS}. We first recall the following compactness result, which can be found in \cite[Theorem XIII.67]{RS} or \cite[Theorem 2.1]{BW}.

\begin{lem}\label{2:lem1}
Suppose $V_i\in L_{loc}^\infty(\R^2)$ satisfies
$\lim_{|x|\to\infty}V_i(x)=\infty$, where $i=1,\, 2$.
Then
the embedding $\mathcal{X}=\h_1\times \h_2\hookrightarrow
L^{q}(\R^2)\times L^{q}(\R^2)$ is compact if $2\leq q<\infty$. \qed
\end{lem}


\noindent\textbf{Proof of Theorem \ref{thm1}:}   We first prove that
(\ref{eq1.3}) admits at least one minimizer if
$0\leq a_i< a^*:=\|Q\|^2_2$, where $i=1,\, 2$.
For any $(u_1,u_2)\in \mathcal{X}$ satisfying $\|u_1\|_2^2=\|u_2\|_2^2=1$, it
 follows from the Gagliardo-Nirenberg inequality (\ref{GNineq}) that
\begin{eqnarray}\label{2:min}
  E(u_1,u_2)\geq\sum_{i=1}^2\int_{\R ^2}\Big[\Big(1-\frac{a_i}{\|Q\|_2^2}\Big)|\nabla
  u_i|^2+V_i(x)|u_i|^2\Big]dx+\beta\int_{\R ^2} u_1^2u_2^2dx.
\end{eqnarray}
Let $\{(u_{1,n},u_{2,n})\}\subset
\mathcal{X}$ be a minimizing sequence of (\ref{eq1.3}), i.e.,
$$\|u_{1,n}\|_2^2=\|u_{2,n}\|_2^2=1\,\ \text{and}\ \
\lim_{n\to\infty}E(u_{1,n},u_{2,n})=e(a_1,a_2).$$
Because of
(\ref{2:min}), we see that $\{(u_{1,n},u_{2,n})\}$ is
bounded in $\mathcal{X}$. By  Lemma
\ref{2:lem1}, there exists  a subsequence of $\{(u_{1,n},u_{2,n})\}$  and $(u_1,u_2)\in
\mathcal{X}$ such that
\begin{equation*}\begin{split}&(u_{1,n},u_{2,n})\overset{n}\rightharpoonup(u_1,u_2)\quad\text{weakly
in}\  \mathcal{X},\\&
(u_{1,n},u_{2,n})\overset{n}\to(u_1,u_2)\quad\text{strongly in}\
L^q(\R^2)\times L^q(\R^2),\ \ \text{where}\,\ 2\leq
q<\infty.\end{split}\end{equation*}
Then, $\|u_1\|_2^2=\|u_2\|_2^2=1$ and $
E(u_1,u_2)=e(a_1,a_2)$ by the weak lower semicontinuity. This proves the existence of minimizers for
all $0\leq a_i< a^* $.

We next prove that if either   $a_1=a_2=a^*$ or  $a_i>a^*$ for $i=1\ \text{or}\ 2$, then  (\ref{eq1.3}) does not admit any minimizer. Let $\varphi(x)\in
C^\infty_0(\R^2)$ be a nonnegative  function such that $\varphi(x)=1$ if $|x|\leq1$
 and $\varphi(x)=0$ if $|x|\geq2$. For any $\bar x_1,\bar x_2\in\R^2$,
 $\tau>0$ and $R>0$, set for $i=1,\, 2$,
 \begin{equation}\label{2:trial}
\phi_i(x)=A_{i,\tau R}\frac{\tau}{\|Q\|_2}\varphi\Big(\frac{x-\bar
x_i+(-1)^iC_0\frac{\ln\tau}{\tau}\vec{n}}{R}\Big)Q\Big(\tau\big[x-\bar
x_i+(-1)^iC_0\frac{\ln\tau}{\tau}\vec{n}\big]\Big),
 \end{equation}
where $A_{i,\tau R}>0$  is chosen such that
$\int_{\R^2}\phi_i^2dx=1$, $\vec{n}\in \R^2$ is a unit vector and
$C_0>1$ is sufficiently large. By
scaling, we deduce from  the exponential decay of $Q$ in (\ref{4:exp}) that
\begin{equation}\label{2:limt1}
\frac{1}{A_{i,\tau R}^2}=\frac{1}{\|Q\|^2_2}\int_{\R^2}\varphi^2(\frac{x}
{\tau R})Q^2(x)dx=1+O((\tau R)^{-\infty}) \text{ as }\tau R\to\infty.
\end{equation}
Here we use the notation $f(t) = O(t^{-\infty})$ for a function $f$ satisfying $\lim_{t\to \infty} |f(t)| t^s = 0$ for all $s>0$. 

By the exponential decay of $Q$, we
deduce from  (\ref{1:id}) that
\begin{equation}\label{2:limt2}
\begin{split}
&\int_{\R^2}|\nabla \phi_i|^2dx-\frac{a_i}{2}\int_{\R^2} \phi_i^4dx\\
=&\,\frac{\tau^2}{\|Q\|^2_2}\int_{\R^2}|\nabla
Q|^2dx-\frac{a_i\tau^2}{2\|Q\|^4_2}\int_{\R^2} Q^4dx+O\big((\tau R)^{-\infty}\big) \\
=&\,\Big(1-\frac{a_i}{a^*}\Big)\tau^2+O((\tau R)^{-\infty})\ \ \text{as}\,\ \tau R\to\infty,
\  \text{where}\,\ i=1,\,2.
\end{split}
\end{equation}
Moreover, we use (\ref{2:limt1}) to obtain that
\begin{equation}\label{2:limt3}
\begin{split}
&\int_{\R^2}
\phi_1^2(x)\phi_2^2(x)dx\\
\leq&\,\frac{2\tau^4}{\|Q\|^4_2}\int_{\R^2}Q^2\Big(\tau(x-\bar
x_1-C_0\frac{\ln\tau}{\tau}\vec{n})\Big)Q^2\Big(\tau(x-\bar
x_2+C_0\frac{\ln\tau}{\tau}\vec{n})\Big)dx\\
\le&\,\frac{2\tau^2}{\|Q\|^2_2}\int_{\R^2}Q^2\Big(x+\tau(\bar x_2-\bar
x_1)-2C_0(\ln\tau)\vec{n}\Big)Q^2(x)dx,
\end{split}
\end{equation}
since $\|Q\|^2_2>1$, cf. Lemma 1 in \cite{GS}.  Note that
\begin{eqnarray}\label{2:tau}
\big|\tau(\bar x_2-\bar x_1)-2C_0(\ln\tau)\vec{n}\big|\geq 2C_0\ln\tau ,\
\text{if}\ \tau>0 \ \text{is sufficiently large,}\end{eqnarray}
denote
$B_{C_0\ln \tau}=\{|x|<C_0\ln \tau\}$, we then derive from (\ref{2:tau})  and (\ref{4:exp})   that,
\begin{equation}\label{2:limt4}\begin{split}
&\int_{\R^2}Q^2\Big(x+\tau(\bar x_2-\bar
x_1)-2C_0(\ln\tau)\vec{n} \Big)Q^2(x)dx \\
=&\int_{B_{C_0\ln \tau}}Q^2\Big(x+\tau(\bar x_2-\bar
x_1)-2C_0(\ln\tau)\vec{n} \Big)Q^2(x)dx \\
 &+\int_{\R^2\setminus B_{C_0\ln
\tau}}Q^2\Big(x+\tau(\bar x_2-\bar
x_1)-2C_0(\ln\tau)\vec{n} \Big)Q^2(x)dx \\
\leq& e^{-2C_0\ln \tau}\Big[\int_{B_{C_0\ln
\tau}}Q^2(x)dx+\int_{\R^2\setminus B_{C_0\ln
\tau}}Q^2\Big(x+\tau(\bar x_2-\bar
x_1)-2C_0(\ln\tau)\vec{n} \Big)dx\Big] \\
\leq& 2\|Q\|_2^2e^{-2C_0\ln \tau}.
\end{split}\end{equation}
It thus follows from (\ref{2:limt3}) and (\ref{2:limt4}) that for $C_0>1$,
\begin{equation}\label{2:limt5}
\int_{\R^2} \phi_1^2(x)\phi_2^2(x)dx\leq 4\tau^2e^{-2C_0\ln
\tau}=4\tau^{2-2C_0}\to 0 \quad {\rm as}\quad \tau\to\infty .
\end{equation}
On the other hand,  the function $x\mapsto
V_i(x)\varphi^2(\frac{x-\bar
x_i+(-1)^iC_0\frac{\ln\tau}{\tau}\vec{n}}{R})$ is bounded and has
compact support. Also note that  $\bar
x_i-(-1)^iC_0\frac{\ln\tau}{\tau}\vec{n}\to \bar x_i$ as
$\tau\to\infty$.  So, there holds for almost every $\bar x_i\in\R^2$ that
\begin{equation}\label{2:limt6}
\lim_{\tau\to\infty}\int_{\R^2}V_i(x)\phi_i^2(x)dx=V_i(\bar x_i), \text{ where }\ i=1,\,2.
\end{equation}

If $a_1=a_2=a^*$, we then conclude from (\ref{2:limt2}), (\ref{2:limt5}) and
(\ref{2:limt6}) that
\begin{equation*}
e(a^*,a^*)\leq V_1(\bar x_1)+V_2(\bar x_2)
\end{equation*}
holds for almost every $\bar x_1,\bar x_2 \in\R^2$. Taking the
infimum over  $\bar x_1$ and $\bar x_2$ in $\R^2$, we  have
$e(a^*,a^*)\leq0$, which and (\ref{2:min}) yield that $e(a^*,a^*)=0$. In this case, suppose there is a minimizer
$(u_1,u_2)\in\mathcal{X}$. As pointed out in the introduction,  we  can assume $(u_1,u_2)$ to be nonnegative. We  then have
\begin{equation*}
\int_{\R^2}|\nabla u_i|^2dx=\frac{1}{2}\int_{\R^2}u_i^4dx\ \ {\rm
and}\ \ \int_{\R^2}V_i(x)u_i^4dx=0,\ \ \mbox{where}\,\  i=1,\,2.
\end{equation*}
This is a contradiction, since the first equality implies that $u_i$
is equal to (up to  translation and scaling) $Q$, and however the second equality yields
that $u_i$ has compact support.

Suppose now that
 $a_i>a^*$ holds for $i=1\ \text{or}\ 2$.
Without loss of generality, we consider the case where $a_1>a^*$. Choose
a positive function $\eta(x)\in C_0^\infty(\R^2)$ such that $\int_{\R^2}\eta^2(x)dx=1$. Then
\begin{equation*}
\int_{\R^2}\phi_1^2\eta^2dx\leq \sup_{x\in
\R^2}\eta ^2(x)\int_{\R^2}\phi_1^2dx=\sup_{x\in \R^2}\eta ^2(x)<\infty,
\end{equation*}
where $\phi_1$ is chosen as in (\ref{2:trial}). We then derive
from (\ref{2:limt2}) that for $a_1>a^*$,
$$E(\phi_1,\eta)\leq \Big(1-\frac{a_1}{\|Q\|_2^2}\Big)\tau^2+C\to -\infty\ \ {\rm as}\,\tau\to \infty,$$
which therefore implies the nonexistence of minimizers.

We finally prove the stated properties of the GP energy $e(a_1,a_2)$. Note that
(\ref{2:min}) implies  $e(a_1,a_2)>0$ for $0\leq a_1,a_2<a^*$. On the other hand, we deduce from (\ref{2:limt2}), (\ref{2:limt5}) and (\ref{2:limt6}) that $e(a^*,a^*)=0$ and $e(a_1,a_2) =
-\infty$ for either $a_1>a^*$ or $a_2>a^*$. Further,    consider  (\ref{2:limt2}), (\ref{2:limt5}) and
(\ref{2:limt6})  by first taking $(a_1,a_2)\nearrow(a^*,a^*)$ and
then letting $\tau\to \infty$, which then implies that
$\limsup_{(a_1,a_2)\nearrow(a^*,a^*)} e(a_1,a_2)\leq V_1(\bar
x_1)+V_2(\bar x_2)$. By taking the infimum over $\bar x_1$
and $\bar x_2$, we then conclude from above estimates that
$\lim_{(a_1,a_2)\nearrow(a^*,a^*)} e(a_1, a_2) = e(a^*, a^*) = 0$, and the proof is therefore complete.\qed

\section{Mass Concentration: Case of $x_1\not =x_2$}
The main purpose of this section is to complete the proof of Theorem \ref{thm2} on the limit behavior of nonnegative minimizers for (\ref{eq1.3}) as  $(a_1, a_2)\nearrow (a^*,a^*)$. Towards this purpose,  we always assume \begin{equation}\label{eq3.1}
\beta >0 \text{ and }V_i(x)=|x-x_i|^{p_i}\text{ with } p_i>0\text{  and } x_1\neq x_2, \text{ where  }i=1,\,2.
\end{equation}
For convenience, as in the introduction, we denote throughout this section
\begin{equation}\label{3:delta}
\eps_i:=(a^*-a_i)^{\frac{1}{p_i+2}},\,\ i=1,\,2;\ \ \delta_0=\delta|x_1-x_2|,
\end{equation}
where $\delta >0$ is given by (\ref{1:del}). Set
\begin{equation}\label{3:limit}
\mathbf{L}:=\overline{\lim}_{(a_1,a_2)\nearrow (a^*,a^*)}
    \frac{e^{\frac{-\delta_0}{\varepsilon_1}}}{\varepsilon_2^{p_2}}\ge 0.
\end{equation}

\subsection{Energy estimates}
In this subsection we  derive some energy estimates for (\ref{eq1.3}) as  $(a_1, a_2)\nearrow (a^*,a^*)$. We start with the following lemma.

\begin{lem}\label{le3.1}
Suppose  (\ref{eq3.1}) holds, and  let
$(u_{a_1},u_{a_2})$ be a nonnegative minimizer of (\ref{eq1.3}). Then  as $(a_1, a_2)\nearrow (a^*,a^*)$,
\begin{equation}\label{3:est}
\beta \int_{\R^2}u_{a_1}^2u_{a_2}^2dx+e_1(a_1)+e_2(a_2)\leq
e(a_1,a_2)\leq
e_1(a_1)+e_2(a_2)+C_0\big(e^{\frac{-\delta_0}{\varepsilon_1}}+
e^{\frac{-\delta_0}{\varepsilon_2}}\big),
\end{equation}
 where
$e_i(a_i)$ is  defined as in (\ref{1:two}). Moreover, if
$\eps_1\geq \eps_2>0$, then,
\begin{equation} e_1(a_1)\leq E_{a_1}^1(u_{a_1})\leq
e_1(a_1)+C_1e^{\frac{-\delta_0}{\varepsilon_1}} \ \ \text{
as}\,\ (a_1, a_2)\nearrow (a^*,a^*),\label{3:a_1}
\end{equation}
 and
\begin{equation}
 e_2(a_2)\leq E_{a_2}^2(u_{a_2})\leq
 e_2(a_2)+C_2e^{\frac{-\delta_0}{\varepsilon_1}}\ \ \text{
as}\, \ (a_1, a_2)\nearrow (a^*,a^*).\label{3:a_2}
\end{equation}
Here, $C_0$, $C_1$ and $C_2$ are some positive constants and all independent of $a_1$ and $a_2$.
\end{lem}

\noindent{\bf Proof.} Since
\begin{equation}\label{3:equ}
E(u,v)=E_{a_1}^1(u)+E_{a_2}^2(v)+\beta \int_{\R^2}u^2v^2dx
\quad\text{for all}\,\ (u,v)\in \mathcal{X},\end{equation}
and note that  $(u_{a_1},u_{a_2})$ is a minimizer of
(\ref{eq1.3}), we then have
\begin{equation}\label{3:left}
e(a_1,a_2)\geq \beta
\int_{\R^2}u_{a_1}^2u_{a_2}^2dx+e_1(a_1)+e_2(a_2).
\end{equation}
On the other hand,  let $\bar u_i$ be a positive minimizer of
$e_i(a_i)$ ($i=1,\,2$). Since $x_1\neq x_2$, it  follows from (\ref{1:dec})
that  \begin{equation*} \bar u_2(x)\leq
Ce^\frac{-\delta|x_1-x_2|}{2\eps_2}\,\ \text{for}\,\ x\in
B_{\frac{|x_1-x_2|}{2}}(x_1),\quad \bar u_1(x)\leq
Ce^\frac{-\delta|x_1-x_2|}{2\eps_1}\,\ \text{for}\,\ x\in \R^2\setminus
B_{\frac{|x_1-x_2|}{2}}(x_1).
\end{equation*}
Hence,
\begin{equation*}
\begin{split} \int_{\R^2}\bar u_1^2\bar u_2^2dx=&\int_{B_{\frac{|x_1-x_2|}{2}}(x_1)}\bar
u_1^2\bar u_2^2dx+\int_{\R^2\setminus B_{\frac{|x_1-x_2|}{2}}(x_1)
}\bar u_1^2\bar
u_2^2dx\\
\leq &C\Big(e^{\frac{-\delta_0}{\varepsilon_2}} \int_{\R^2}\bar
u_1^2dx+e^{\frac{-\delta_0}{\varepsilon_1}} \int_{\R^2}\bar
u_2^2dx\Big)=C\big(e^{\frac{-\delta_0}{\varepsilon_1}}+
e^{\frac{-\delta_0}{\varepsilon_2}}\big),
\end{split}
\end{equation*}
where $\delta_0=\delta|x_1-x_2|$. This yields that
\begin{equation}\label{3:rig}
e(a_1,a_2)\leq E(\bar u_1,\bar u_2)\leq e_1(a_1)+e_2(a_2)+2\beta
C\big(e^{\frac{-\delta_0}{\varepsilon_1}}+e^{\frac{-\delta_0}{\varepsilon_2}}\big),
\end{equation}
by which we then conclude (\ref{3:est}) in view of (\ref{3:left}).

We next turn to proving (\ref{3:a_1}) and (\ref{3:a_2}). By  the definition of $e_i(\cdot)$, it is clear to have the
lower bounds of (\ref{3:a_1}) and (\ref{3:a_2}), then we need only to get the upper bounds. We first claim
that \begin{equation}\label{3:right}E_{a_1}^1(u_{a_1})\leq
e_1(a_1)+2C_0e^{\frac{-\delta_0}{\varepsilon_1}}\ \ \text{as}\,\ (a_1,a_2)\nearrow
(a^*,a^*).\end{equation}
On the
contrary, suppose (\ref{3:right}) is false. Then there exists a subsequence of $\{a_1\}$, still denoted by
$\{a_1\}$, such that
$$E_{a_1}^1(u_{a_1})\geq e_1(a_1)+Ce^{\frac{-\delta_0}{\varepsilon_1}}  \ \ \text{as}\ \ (a_1,a_2)\nearrow
(a^*,a^*) $$
for some constant $ C>2C_0$. By applying (\ref{3:est}) and (\ref{3:equ}), we  deduce from the above that
$$e_1(a_1)+e_2(a_2)+Ce^{\frac{-\delta_0}{\varepsilon_1}}\leq E_{a_1}^1(u_{a_1})+E_{a_2}^2(u_{a_2})\leq e_1(a_1)+e_2(a_2)+C_0\big(e^{\frac{-\delta_0}{\varepsilon_1}}+
e^{\frac{-\delta_0}{\varepsilon_2}}\big).$$
  Since $\eps_1\geq\eps_2$ implies that
$e^{\frac{-\delta_0}{\varepsilon_1}}\geq
e^{\frac{-\delta_0}{\varepsilon_2}}$, we  conclude that $C\leq 2C_0$, which leads to  a
contradiction. We thus  have (\ref{3:right}), and  (\ref{3:a_1}) is therefore
established. Similarly, one can derive (\ref{3:a_2}) for some $C_2\leq
2C_0$, and we are done.  \qed

 \vspace {.2cm}

%
Motivated by Lemma 4 in \cite{GS}, we next use Lemma \ref{le3.1} to address the following  $L^4(\R^2)$
estimates of minimizers.

\begin{lem}\label{le3.2} Assume (\ref{eq3.1}) holds and $\eps_1\geq\eps_2>0$. Let $(u_{a_1},u_{a_2})$ be a non-negative
minimizer of (\ref{eq1.3}).  Then, there exists a positive constant
$K_1 $, independent of $a_1$ and $a_2$, such that
\begin{equation}\label{eq2.6}
K_1 (a^*-a_1)^{-\frac{2}{p_1+2}}\leq \int_{\mathbb{R}^2}|
u_{a_1}|^4dx\leq \frac{1}{K_1}(a^*-a_1)^{-\frac{2}{p_1+2}} \ \
\text{as}\,\  (a_1,a_2)\nearrow (a^*,a^*).
\end{equation}
Moreover, there exists a positive constant
$K_2 $, independent of $a_1$ and $a_2$, such that
\begin{itemize}
\item[\em(i).] If  the limit $\mathbf{L}$ in (\ref{3:limit}) satisfies $0\le \mathbf{L}<\infty$,
 then
\begin{equation}\label{3:finity}
 K_2 (a^*-a_2)^{-\frac{2}{p_2+2}}\leq \int_{\mathbb{R}^2}|
u_{a_2}|^4dx\leq \frac{1}{K_2}(a^*-a_2)^{-\frac{2}{p_2+2}} \ \
\text{as}\, \ (a_1, a_2)\nearrow (a^*,a^*).
\end{equation}
\item[\em(ii).] If   $\mathbf{L}=\infty$,
then
\begin{equation}\label{3:infinity}
 K_2
\big(e^\frac{-\delta_0}{\varepsilon_1}\big)^{-\frac{2}{p_2}}\leq
\int_{\mathbb{R}^2}| u_{a_2}|^4dx\leq
\frac{e^\frac{-\delta_0}{\varepsilon_1}}{K_2\varepsilon_2^{p_2+2}}
\ \ \text{as}\,\   (a_1, a_2)\nearrow (a^*,a^*).
\end{equation}
\end{itemize}
\end{lem}

\noindent{\bf Proof.} (a).\, By (\ref{1:e}) and (\ref{3:a_1}), there exists a constant $C>0$, independent of $a_1$ and $a_2$, such that
\begin{equation}\label{3:L1}
E_{a_1}^1(u_{a_1})\leq Ce_1(a_1).
\end{equation}
On the other hand, using (\ref{GNineq}), we
obtain that
\begin{equation}\label{3:L}
E_{a_1}^1(u_{a_1})\geq\frac{a^*-a_1}{2}\int_{\mathbb{R}^2}|u_{a_1}(x)|^4dx,
\end{equation}
and the upper bound of (\ref{eq2.6}) hence follows from (\ref{1:e}) and (\ref{3:L1}).

To prove the lower bound of (\ref{eq2.6}), we take $0<b<a_1<a^*$ and then have
\begin{equation*}
e_1(b)\leq
E_{a_1}^1(u_{a_1})+\frac{a_1-b}{2}\int_{\mathbb{R}^2}u_{a_1}^4(x)dx.
\end{equation*}
Together with  (\ref{1:e}) and (\ref{3:L1}), this implies that
there exist  $C_1,C_2>0$ such that
\begin{equation}\label{eq2.7}
\frac{1}{2}\int_{\mathbb{R}^2}|u_{a_1}(x)|^4dx\geq
\frac{e_1(b)-Ce_1(a_1)}{a_1-b}\geq
\frac{C_1(a^*-b)^\frac{p_1}{p_1+2}-C_2(a^*-a_1)^\frac{p_1}{p_1+2}}{a_1-b}.
\end{equation}
Take $b=a_1-C(a^*-a_1)$, where $C>0$ is large so that
$C_1C^\frac{p_1}{p_1+2}>2C_2$. We then derive from (\ref{eq2.7})
that
\begin{equation*}
\int_{\mathbb{R}^2}|u_{a_1}(x)|^4dx\geq
\frac{2C_2}{C}(a^*-a_1)^\frac{p_1}{p_1+2},
\end{equation*}
which then implies the lower bound of (\ref{eq2.6}).

(b).\, If  $0\le\mathbf{L}<\infty$,   it follows from  (\ref{1:e}) and (\ref{3:a_2})
that there exists a constant $C>0$, independent of $a_1$ and $a_2$, such that
\[E^2_{a_2}(u_{a_2})\leq Ce_2(a_2)\quad \mbox{for}\quad 0\le a_2\le a^*.\]
Further, a similar
proof of (\ref{eq2.6}) then implies (\ref{3:finity}).

(c).\, If  $\mathbf{L}=+\infty$,  by (\ref{1:e}) and (\ref{3:a_2}) we then have
$$e_2(a_2)\leq E^2_{a_2}(u_{a_2})\leq C_3e^\frac{-\delta_0}{\eps_1}.$$
On the other hand, similar to  (\ref{3:L}), we have
\begin{equation*}
E_{a_2}^2(u_{a_2})\geq\frac{a^*-a_2}{2}\int_{\mathbb{R}^2}|u_{a_2}(x)|^4dx,
\end{equation*}
The above inequalities  give the upper bound of
(\ref{3:infinity}). To obtain the lower bound, we take
$b=a_2-(Me^\frac{-\delta_0}{\eps_1})^\frac{p_2+2}{p_2}$, where $M>0$
is sufficiently large, so that
$$(a^*-b)^\frac{p_2}{p_2+2}=\big[a^*-a_2+(Me^\frac{-\delta_0}{\eps_1}
)^\frac{p_2+2}{p_2}\big]^\frac{p_2}{p_2+2}\geq Me^\frac{-\delta_0}{\eps_1}. $$
Similar to (\ref{eq2.7}), we then have
\begin{equation*}
\frac{1}{2}\int_{\mathbb{R}^2}|u_{a_2}(x)|^4dx\geq
\frac{e_2(b)-C_3e^\frac{-\delta_0}{\eps_1}}{a_2-b}\geq
\frac{C_1(a^*-b)^\frac{p_2}{p_2+2}-C_3e^\frac{-\delta_0}{\eps_1}}{a_2-b}\geq
C\big(e^\frac{-\delta_0}{\varepsilon_1}\big)^{-\frac{2}{p_2}},
\end{equation*}
which therefore implies the lower bound of (\ref{3:infinity}). \qed

\subsection{Proof of Theorem \ref{thm2}}

This subsection is devoted to the proof of Theorem \ref{thm2}.  As before, we still
denote $(u_{a_1},u_{a_2})$  a non-negative minimizer of
(\ref{eq1.3}). We first establish the following theorem, which gives (i) and (ii) of Theorem \ref{thm2}.

\begin{thm}\label{thm3.1} Assume that   (\ref{eq3.1}) holds, $\eps_1\geq\eps_2>0$, and  the limit $\mathbf{L}$ in (\ref{3:limit}) satisfies $0\le \mathbf{L}<\infty$. Then each $u_{a_i}$ has a unique global maximum point $x_{a_i}$ which approaches $x_i$ as $(a_1,a_2)\nearrow (a^*,a^*)$, where $i=1,\, 2$. Moreover, we have
\begin{itemize}
\item[\em(i).] $u_{a_1}$ satisfies \begin{equation}\label{3:conv1} \lim_{(a_1, a_2)\nearrow (a^*,a^*)} \varepsilon_1
u_{a_1}(\varepsilon_1
x+x_{a_1})=\frac{\lambda_1}{\|Q\|_{2}}Q(\lambda_1x) \text{ strongly in $H^1(\R^2)$},
 \end{equation}where $x_{a_1}$ satisfies
\begin{equation}\label{3:lm1}
\frac{|x_{a_1}-x_1|}{\varepsilon_1}\to 0 \ \ \text{as}\,\
(a_1,a_2)\nearrow (a^*,a^*),
\end{equation}
 and
\begin{equation}\label{3:lm2}
  \lambda_1 = \Big(
\frac{p_1}{2}\inte | x|^{p_1}Q^2(x)dx\Big)^\frac{1}{p_1+2}.
\end{equation}

\item[\em(ii).]  $u_{a_2}$ satisfies
\begin{equation}\label{3a:lm1} \lim_{(a_1, a_2)\nearrow (a^*,a^*)} \varepsilon_2
u_{a_2}(\varepsilon_2
x+x_{a_2})=\frac{\lambda_2}{\|Q\|_{2}}Q(\lambda_2x) \text{  in $H^1(\R^2)$ for some $\lambda_2>0$.}
 \end{equation}
Specially, if $\mathbf{L}=0$, then (\ref{3a:lm1}) holds for
\begin{equation}
\frac{|x_{a_2}-x_2|}{\varepsilon_2}\to 0 \ \ \text{as}\,\
(a_1,a_2)\nearrow (a^*,a^*),\label{3:lm3}
\end{equation}
and \begin{equation}\lambda_2 = \Big( \frac{p_2}{2}\inte |
x|^{p_2}Q^2(x)dx\Big)^\frac{1}{p_2+2}.\label{3:lm4}
\end{equation}
\end{itemize}
\end{thm}

\noindent{\bf Proof.} We establish this theorem by three steps:
 \vspace {.2cm}

\noindent{\em Step 1: The proofs of (\ref{3:conv1}) and (\ref{3a:lm1}).} We first note
from (\ref{3:a_1}), (\ref{3:a_2}),
(\ref{eq2.6}) and (\ref{3:finity})  that there
exist positive constants $C_1$, $C_2$ and $C_3$, independent of
$a_1$ and $a_2$, such that
\begin{equation}\label{eq2.8}
\int_{\mathbb{R}^2}V_i(x)|u_{a_i}(x)|^2dx\leq C_1\varepsilon_i^{p_i}
\ \ \text{as}\,\   (a_1,a_2)\nearrow (a^*,a^*),
\end{equation}
and \begin{equation}\label{eq2.9}
C_2\varepsilon_i^{-2}\leq\int_{\mathbb{R}^2}| u_{a_i}(x)|^4dx\leq
C_3\varepsilon_i^{-2}\ \ \text{as}\,\   (a_1,a_2)\nearrow (a^*,a^*),\,\ i=1,\,2,
\end{equation}
provided that $0\le \mathbf{L}<\infty$.
Since Lemma \ref{le3.1} gives that
\begin{equation*}
\begin{split}
0&\leq\int_{\mathbb{R}^2}|\nabla
u_{a_i}(x)|^2dx-\frac{a_i}{2}\int_{\mathbb{R}^2}|
u_{a_i}(x)|^4dx\\
&\leq
e_i(a_i)+C_1e^{\frac{-\delta_0}{\varepsilon_1}}\le C\varepsilon_i^{p_i}\ \
\text{ as }\,\ (a_1,a_2)\nearrow (a^*,a^*),
\end{split}
\end{equation*}
there exists a constant $m>0$ such that
\begin{equation}\label{eq2.10}
0<m\varepsilon_i ^{-2}<\int_{\mathbb{R}^2}| \nabla
u_{a_i}(x)|^2dx<\frac{1}{m}\varepsilon_i ^{-2}\ \ \text{ as }\,\
(a_1,a_2)\nearrow (a^*,a^*),\,\ i=1,\,2.
\end{equation}

\noindent \textbf{Claim 1:} Each $u_{a_i}$ ($i=1,2$) has at least one local maximum point  $x_{a_i}$, and there exists $\eta>0$ such that the function
\begin{equation}\label{3:W}\bar{w}_{a_i}=\varepsilon_iu_{a_i}(\varepsilon_i x+x_{a_i}),
\end{equation}
satisfies \begin{equation}\label{eq2.14}
\liminf_{(a_1,a_2)\nearrow (a^*,a^*)}\int_{B_{2}(0)}|\bar{w}_{a_i}|^2dx\geq\eta>0,\,\ i=1,\,2.
\end{equation}

Indeed, since $(u_{a_1},u_{a_2})$ is a nonnegative minimizer of (\ref{eq1.3}),
it satisfies the Euler-Lagrange system
\begin{equation}
\begin{cases}
 -\Delta u_{a_1}+V_1(x)u_{a_1}=\mu_{a_1} u_{a_1}+a_1
u_{a_1}^3-\beta u_{a_2}^2u_{a_1}\ \ \mbox{in}\,\ \R^2,\label{3:ua}\\[2mm]
-\Delta u_{a_2}+V_2(x)u_{a_2}=\mu_{a_2} u_{a_2}+a_2 u_{a_2}^3-\beta
u_{a_1}^2u_{a_2}\ \ \mbox{in}\,\ \R^2,
\end{cases}
\end{equation}
where $(\mu_{a_1},\mu_{a_2})$  is a suitable Lagrange multiplier satisfying
\begin{equation}\label{eq3.28}
\mu_{a_i}=E^i_{a_i}(u_{a_i})-\frac{a_i}{2}\int_{\mathbb{R}^2}|u_{a_i}|^4dx
+\beta\int_{\R^2}u_{a_1}^2u_{a_2}^2dx,\,\ i=1,\,2.
\end{equation}
One can use the comparison principle as in \cite{KW} to deduce from (\ref{3:ua}) that $u_{a_i}$ decays exponentially to zero  at infinity.
This implies that each $u_{a_i}$ has at least one local maximum point, which is denoted by $x_{a_i}$. Letting $\bar w_{a_i}$ be defined by (\ref{3:W}),
it then follows from (\ref{3:ua}) that
{\small \begin{equation}\label{3:222}\begin{cases} -\Delta \bar{w}_{a_1}+\varepsilon_1^2
V_1(\varepsilon_1 x+x_{a_1})\bar{w}_{a_1}=\mu_{a_1}\varepsilon_1^2
\bar{w}_{a_1}+a_1
\bar{w}_{a_1}^{3}-\beta(\frac{\eps_1}{\eps_2})^2\bar
w_{a_2}^2(\frac{\varepsilon_1
x+x_{a_1}-x_{a_2}}{\eps_2})\bar w_{a_1}\ \ \mbox{in}\,\ \R^2, \\[2mm]
-\Delta \bar{w}_{a_2}+\varepsilon_2^2 V_2(\varepsilon_2
x+x_{a_2})\bar{w}_{a_2}=\mu_{a_2}\varepsilon_2^2 \bar{w}_{a_2}+a_2
\bar{w}_{a_2}^{3}-\beta(\frac{\eps_2}{\eps_1})^2\bar
w_{a_1}^2(\frac{\varepsilon_2 x+x_{a_2}-x_{a_1}}{\eps_1})\bar
w_{a_2}\ \ \mbox{in}\,\ \R^2.
\end{cases}\end{equation}}
Moreover, applying Lemma \ref{le3.1}, (\ref{eq2.9}) and (\ref{eq3.28}),   we see that there
exist two positive constants $C_1$ and $C_2$, independent of $a_1$ and $a_2$,
such that
$$-C_2<\varepsilon_i^2\mu_{a_i} <-C_1<0\ \ \text{as}\,\   (a_1,a_2)\nearrow (a^*,a^*), \,\ i=1,\,2.$$
Therefore, by passing to a subsequence if necessary, we may assume
that
\begin{equation}\label{eq3.30}
\lim_{(a_1,a_2)\nearrow(a^*,a^*)}\eps_i^2\mu_{a_i}=-\lambda _i^2\ \  \text{ for some }\,\ \lambda _i>0, \,\ i=1,2.
\end{equation}
Since $\beta>0$, we then obtain from  (\ref{3:222}) that
\begin{equation}\label{bound}
\bar{w}_{a_i}(0)\geq \Big(\frac{-\mu_{a_i}\varepsilon ^2_i}{a
_i}\Big)^\frac{1}{2}\geq
\Big(\frac{\lambda_i^2}{2a^*}\Big)^\frac{1}{2}\ \ \text{as}\,\
(a_1,a_2)\nearrow (a^*,a^*).
\end{equation}
By (\ref{3:222}), we also have
\begin{equation*}
-\Delta \bar w_{a_i}-c_i(x)\bar w_{a_i}\leq 0\ \ \mbox{in}\,\ \R^2,\end{equation*}
where $c_i(x)=a_i \bar w_{a_i}^2(x)$ ($i=1,2)$. Applying De Giorgi-Nash-Moser
theory, we thus have
$$\max_{B_1(\xi)}  \bar w_{a_i}\leq
C\Big(\int_{B_2(\xi)}|  \bar w_{a_i}|^2 dx\Big)^\frac{1}{2},$$
where $\xi$ is an arbitrary point in $\mathbb{R}^2$, and $C$ is a
constant depending only on the bound of
$\|  \bar w_{a_i}\|_{L^4(B_2(\xi))}$.  Taking $\xi=0$, we then obtain from (\ref{bound}) that
$$\liminf_{(a_1,a_2)\nearrow (a^*,a^*)}\int_{B_{2}(0)}|\bar{w}_{a_i}|^2dx\geq \frac{\min\{\lambda_1^2,\lambda_2^2\}}{2C^2 a^*}:=\eta>0,\,$$
i.e., (\ref{eq2.14}) holds, and the claim is thus established.

We next prove that \begin{equation}\label{eq3.32}
x_{a_i}\to x_i \ \ \text{as}\,\   (a_1,a_2)\nearrow (a^*,a^*), \,\ i=1,2.
\end{equation}
On the contrary, if (\ref{eq3.32}) is incorrect for $i=1$ or $2$, then there exist a subsequence, still denoted by $\{a_i\}$, of $\{a_i\}$ and a constant $C>0$ such that
$$V_i(x_{a_i})\geq C>0 \ \ \text{as}\,\   (a_1,a_2)\nearrow (a^*,a^*), \ \ \text{where}\  \ i=1\ \text{or }2. $$
Using  Fatou's Lemma, we thus obtain from  (\ref{3:W}) and (\ref{eq2.14}) that
\begin{equation*}
\begin{split}
&\lim_{(a_1,a_2)\nearrow (a^*,a^*)}\int_{\mathbb{R}^2}V_i(x)|u_{a_i}(x)|^2dx\\
&\quad=\lim_{(a_1,a_2)\nearrow (a^*,a^*)}\int_{\mathbb{R}^2}V_i(\eps_ix+x_{a_i})|\bar w_{a_i}(x)|^2dx
\geq \frac{C\eta}{2}>0,\
\ \ i=1\ \text{or }2,
\end{split}
\end{equation*}
this contradicts (\ref{eq2.8}). Thus (\ref{eq3.32}) holds.

From (\ref{eq2.9}) and (\ref{eq2.10}) we see that $\{\bar w_{a_i}\}$ is bounded in $H^1(\R^2)$.
Thus, there exists a subsequence and $0\leq \bar w_i(x)\in H^1(\R^2)$ such that
 $\bar{w}_{a_i}\rightharpoonup
\bar{w}_i$ in $H^1(\mathbb{R}^2)$, and it follows from (\ref{3:222}), (\ref{eq3.30}) and  (\ref{eq3.32}) that $\bar w_i$ satisfies
\begin{equation}\label{eq2.20}
-\Delta w(x)=-\lambda_i^2 w(x)+a^* w^{3}(x)\ \ \text {in} \,\ \mathbb{R}^2,\ \ \mbox{where} \,\ i=1,\,2.
\end{equation}
Moreover, we have $\bar w_i(x)\not\equiv0$ by (\ref{eq2.14}), and the strong maximum
principle then yields that $\bar{w}_i(x)>0$ in $\R^2$. Since the origin
is a critical point of $\bar{w}_{a_i}$, it is also a critical point
of $\bar{w}_i$.  We therefore conclude from the uniqueness (up to
translations) of positive radial solutions for (\ref{Kwong}) that
$\bar{w}_i$ is spherically symmetric about the origin, and
\begin{equation}\label{eq2.350}
\bar w_i=\frac{\lambda_i}{\|Q\|_2}Q(\lambda_i|x|) \ \ \text{for
some}\,\ \lambda_i >0,
\end{equation}
where $\|\bar w_i\|_2^2=1$ ($i=1,\,2$). By the norm preservation we further
conclude that $\bar w_{a_i}\to\bar w_i$ strongly in
$L^2(\mathbb{R}^2)$. Moreover, it follows from (\ref{3:222}) and (\ref{eq2.20}) that
 $$\lim_{(a_1,a_2)\nearrow(a^*,a^*)}\bar w_{a_i}=\bar w_i=\frac{\lambda_i}{\|Q\|_2}Q(\lambda_i|x|) \  \text{ in }H^1(\mathbb{R}^2), \,\ i=1,2,$$
and therefore (\ref{3:conv1}) and (\ref{3a:lm1}) are proved.

\vskip 0.1truein

\noindent{\em Step 2: Uniqueness of global maximal points.} Since $\bar
w_{a_i}(x)\rightarrow \bar w_i(x)$ strongly in $H^1(\mathbb{R}^2)$,
similar to Step 1  by applying the comparison principle and De Giorgi-Nash-Moser theory, one can derive that for  $i=1,\,2$,
\begin{equation}\label{3:exp} \bar w_{a_i}(x)\leq C
e^{-\frac{\lambda_i}{2}|x|} \ \ \text{for} \ \ |x|>R \ \
\text{as}\ \ (a_1,a_2)\nearrow (a^*,a^*),
\end{equation}
and
\begin{equation}\label{3:lin}
\|\bar w_{a_i}(x)\|_{L^\infty(\R^2)}\leq C<\infty \ \
\text{as}\,\
(a_1,a_2)\nearrow (a^*,a^*),
\end{equation}
where $ C>0$ is independent of $a_1$ and $a_2$.

\vskip 0.1truein
\noindent\textbf{Claim 2:}
 $\bar{w}_{a_i}(x)\rightarrow \bar{w}_i(x)$ in
 $C^{2,\alpha}_{loc}(\mathbb{R}^2)$ for some $\alpha>0$, where $i=1,\,2$.
Let $\Omega_1\subset\subset\Omega\subset\subset\R^2$. 
Rewrite (\ref{3:222}) as
\begin{equation}\label{3:hold}
-\Delta \bar{w}_{a_i}(x)=g_{a_i}(x)\ \ \mbox{in}\,\ \R^2,\,\ i=1,\, 2,
\end{equation}
where
$$g_{a_1}(x)=-\varepsilon_1^2 V_1(\varepsilon_1
x+x_{a_1})\bar{w}_{a_1}+\mu_{a_1}\varepsilon_1^2 \bar{w}_{a_1}+a_1
\bar{w}_{a_1}^{3}-\beta \Big(\frac{\eps_1}{\eps_2}\Big)^2\bar w_{a_2}^2\Big(\frac{\varepsilon_1
x+x_{a_1}-x_{a_2}}{\eps_2}\Big)\bar w_{a_1},$$
and
$$g_{a_2}(x)=-\varepsilon_2^2 V_2(\varepsilon_2
x+x_{a_2})\bar{w}_{a_2}+\mu_{a_2}\varepsilon_2^2 \bar{w}_{a_2}+a_2
\bar{w}_{a_2}^{3}-\beta \Big(\frac{\eps_2}{\eps_1}\Big)^2\bar w_{a_1}^2\Big(\frac{\varepsilon_2
x+x_{a_2}-x_{a_1}}{\eps_1}\Big)\bar w_{a_2}.$$
Since $\eps_1\geq\eps_2>0$, we obtain from (\ref{3:exp}) and (\ref{3:lin}) that
$$\|g_{a_2}(x)\|_{L^\infty(\R^2)}\leq C<\infty
\ \ \text{uniformly as}\ \ (a_1,a_2)\nearrow (a^*,a^*).$$
Then,  the $L^p$ theory of \cite{GT} yields that $\bar w_{a_2}(x)\in W^{2,q}_{loc}(\R^2)$
for any $q>2$. It thus follows from Sobolev's imbedding theorem that
$\bar w_{a_2}(x)\in C^{1,\alpha}_{loc}(\R^2)$ for some $\alpha\in(0,1)$. Further, we deduce from Theorem 8.32 in \cite{GT} that,
\begin{equation}\label{3:hold1}
\|\bar w_{a_2}\|_{C^{1,\alpha}(\R^2)}\leq C\Big(\|\bar w_{a_2}\|_{L^\infty(\R^2)}+\|\bar g_{a_2}\|_{L^\infty(\R^2)}\Big)\leq C.
\end{equation}

On the other hand,  since $x_1\neq
x_2$, we see from (\ref{eq3.32}) that as $(a_1,a_2)\nearrow (a^*,a^*)$,
$$\frac{|\eps_1x+x_{a_1}-x_{a_2}|}{\eps_2}\geq \frac{|x_{1}-x_{2}|}{2\eps_2}\to \infty \ \ \text{for all}\ \ x\in\Omega.$$
This estimate and (\ref{3:exp}) imply that, for any $\gamma\ge2$,
\begin{equation}\label{3:lii}\Big(\frac{\eps_1}{\eps_2}\Big)^\gamma\bar w_{a_2}^2\Big(\frac{\varepsilon_1
x+x_{a_1}-x_{a_2}}{\eps_2}\Big)\leq
C\Big(\frac{\eps_1}{\eps_2}\Big)^\gamma e^{-\frac{\lambda _2|x_1-x_2|}{\eps_2}}\to 0\: \
\text{for all}\:\ x\in\Omega.\end{equation}
By (\ref{3:lin}) and
(\ref{3:lii}), there exists $C>0$,  depending  only on $\Omega$, such that
$$\|g_{a_1}(x)\|_{L^\infty(\Omega)}\leq C<\infty \text{ uniformly as $(a_1,a_2)\nearrow (a^*,a^*)$}.$$
Then,  similar to the arguments of analyzing $ \bar w_{a_2}(x)$, we deduce from the  $L^p$ theory and  Theorem 8.32 in \cite{GT} that
$\bar w_{a_1}(x)\in C^{1,\alpha}(\Omega)$ for some $\alpha\in(0,1)$ and
\begin{equation}\label{3:hold3}
\|\bar w_{a_1}\|_{C^{1,\alpha}(\Omega_1)}\leq C\big(\|\bar w_{a_1}\|_{L^\infty(\Omega)}+\| g_{a_1}\|_{L^\infty(\Omega)}\big)\leq C<\infty.
\end{equation}
Meanwhile, (\ref{3:hold1}) and (\ref{3:lii}) yield that
$$\Big|\big(\frac{\eps_1}{\eps_2}\big)^2\nabla _x\bar w_{a_2}^2\big(\frac{\varepsilon_1
x+x_{a_1}-x_{a_2}}{\eps_2}\big)\Big|=2\Big|\big(\frac{\eps_1}{\eps_2}\big)^3\big(\bar w_{a_2}\nabla _y\bar w_{a_2}\big)\big(\frac{\varepsilon_1
x+x_{a_1}-x_{a_2}}{\eps_2}\big)\Big|\leq C \ \ \text{in}\, \ \Omega,$$
where $y:=\frac{\varepsilon_1 x+x_{a_1}-x_{a_2}}{\eps_2}$.
This estimate and (\ref{3:hold3}) imply that
$$\|g_{a_1}(x)\|_{ C^1(\Omega_1)}\le C<\infty\ \ \text{uniformly as}\ \ (a_1,a_2)\nearrow (a^*,a^*).$$
Then, in view of (\ref{3:hold}), Schauder estimates yield that $\bar w_{a_1}(x)$ is bounded uniformly in $C^{2,\alpha}(\Omega_1)$ for some $\alpha\in(0,1)$, which  implies  that Claim 2  holds for $i=1$.

To prove that Claim 2 holds also for $i=2$, we first prove that
\begin{equation}\label{3:gradient}
|\nabla \bar w_{a_1}(x)|\leq C\frac{\eps_1}{\eps_2}e^{-\frac{\lambda_1}{4}|x|}\ \ \text{for} \ \ |x|>R+2 \ \
\text{as}\ \ (a_1,a_2)\nearrow (a^*,a^*),
\end{equation}
where the radius $R>0$ is as in  (\ref{3:exp}).
 Indeed, for any fixed $x\in\R^2$ with $|x|>R+2$, let $y\in B_2(0)$ and define
 \begin{equation*}
\hat w_{a_1}(y):= \bar w_{a_1}\big(x+\frac{\eps_2}{\eps_1}y\big).
\end{equation*}
Then, $\hat w_{a_1}(y)$ satisfies
\begin{equation}\label{3:hold5}
\begin{split}
-\Delta \hat w_{a_1}(y)=&\Big(\frac{\eps_2}{\eps_1}\Big)^2\Big[-\varepsilon_1^2 V_1(\eps_1x+\eps_2 y+x_{a_1})\hat{w}_{a_1}(y)+\mu_{a_1}\varepsilon_1^2 \hat{w}_{a_1}(y)+a_1
\hat{w}_{a_1}^{3}(y)\Big]\\
&-\beta \bar w_{a_2}^2\Big(\frac{\varepsilon_1
x+\eps_2y+x_{a_1}-x_{a_2}}{\eps_2}\Big)\hat w_{a_1}(y):=\hat g_{a_1}(y)\ \ \mbox{in}\,\ \R^2.
\end{split}
\end{equation}
By (\ref{3:exp}),
$$\hat w_{a_1}(y)\leq Ce^{-\frac{\lambda_1}{2}|x|}\ \ \text{for}\,\ y\in B_2(0),$$
which and (\ref{3:lin}) thus imply that
$$|\hat g_{a_1}(y)|\leq Ce^{-\frac{\lambda_1}{4}|x|}\ \ \text{for}\,\  y\in B_2(0).$$
Hence, using (\ref{3:hold5}) and employing  Theorem 8.32 in \cite{GT} again, we obtain that
$$\|\nabla\hat w_{a_1}(y)\|_{L^\infty(B_1(0))}\leq Ce^{-\frac{\lambda_1}{4}|x|}\ \ \text{for}\,\  y\in B_2(0),$$
which therefore implies that (\ref{3:gradient}) holds.

We now derive from (\ref{3:hold1}), (\ref{3:hold3}) and  (\ref{3:gradient}) that
$$\|g_{a_2}(x)\|_{ C^1(\Omega_1)}<C\ \ \text{uniformly as}\,\ (a_1,a_2)\nearrow (a^*,a^*).$$
Therefore, by Schauder estimates, $\bar w_{a_2}(x)$ is bounded uniformly in $C^{2,\alpha}(\Omega_1)$, and hence,
\begin{equation*}
\bar{w}_{a_2}(x)\rightarrow \bar{w}_2(x)\ \ \text{in}\,\ C^{2,\alpha}_{loc}(\mathbb{R}^2)\ \ \text{for some}\,\ \alpha\in(0,1),\end{equation*}
which implies that Claim 2 holds also for $i=2$.
 \vspace {.2cm}

Using the above claim,
similar to the arguments of \cite{Wang}, one can deduce  from Lemma 4.2 in \cite{NT} that
for $(a_1,a_2)\nearrow (a^*,a^*)$, $\bar{w}_{a_i}$ has no critical points other than the
origin. This  gives the uniqueness of local maximum points
for $\bar{w}_{a_i}(x)$, which   also indicates  the {\em uniqueness} of global maximum point of $u_{a_i}$ in view of (\ref{3:W}).

\vskip 0.1truein

\noindent{\em Step 3: The proofs of (\ref{3:lm1}), (\ref{3:lm2}), (\ref{3:lm3}) and (\ref{3:lm4}).}  It follows from
(\ref{3:a_1}) that \begin{equation*}
E^1_{a_1}(u_{a_1})=e_1(a_1)+o(1)e_1(a_1) \ \ \text{as}\,\
(a_1,a_2)\nearrow (a^*,a^*),
\end{equation*}
which together with (\ref{GNineq}) and  (\ref{3:W}) yields  that
\begin{align}\label{3:eqa}
&e_1(a_1)[1+o(1)]= E^1_{a_1}(u_{a_1}) \nonumber
\\  \geq &\frac{\eps_1^{p_1}}2 \inte \bar{w}^4_{a_1} (x) dx + \inte
\big|\eps_1 x+x_{a_1}-x_1\big|^{p_1} \bar{w}_{a_1}^2 (x) dx\nonumber\\
=&\frac{\eps_1^{p_1}}2 \inte \bar{w}^4_{a_1} (x) dx + \eps_1^{p_1}\inte
\big|x+\frac{x_{a_1}-x_1}{\eps_1}\big|^{p_1} \bar{w}_{a_1}^2 (x) dx
\end{align}
as $(a_1,a_2)\nearrow (a^*,a^*)$, where $x_{a_1}$ is the unique global maximum  point of $u_{a_1}$ satisfying (\ref{eq3.32}).
Utilizing (\ref{eq2.14}) and (\ref{1:e}), one can  deduce from (\ref{3:eqa}) that
$\frac{|x_{a_1}-x_1|}{\varepsilon_1}$ is bounded  as
$(a_1,a_2)\nearrow (a^*,a^*)$.
Therefore, we can extract a  subsequence such that
\begin{equation*}
\frac{x_{a_1}-x_1}{\varepsilon_1} \to z_1 \ \ \text{as}\,\   (a_1,a_2)\nearrow (a^*,a^*)\, \text{ for some $z_1\in\R^2$. }
\end{equation*}
Since $Q$ is a radial
decreasing function and decays exponentially as $|x|\to\infty$, we
then deduce from (\ref{eq2.350}) that
\begin{align}\label{str}
 &\liminf_{(a_1,a_2)\nearrow (a^*,a^*)}\frac{1}{\eps_1^{p_1}} \inte \big|\eps_1
 x+x_{a_1}-x_1\big|^{p_1} \bar w_{a_1}^2 (x) dx \nonumber\\
 &=\inte |x+z_1|^{p_1} \bar w_0^2(x)dx\geq \inte |x|^{p_1} \bar w_0^2(x)dx,
\end{align}
where ``=" holds in the last inequality  if and only if $z_1=(0,0)$. We hence infer
from (\ref{1:id}), (\ref{1:e}), (\ref{3:eqa}) and (\ref{str})  that
\begin{equation}\label{mb}
\begin{split} \liminf_{(a_1,a_2)\nearrow (a^*,a^*)}\frac{e_1(a_1)}{\eps_1^{p_1}} &\geq
\frac 12 \|\bar w_0\|_4^4 +\inte |x|^{p_1} \bar w_0^2(x)dx \\
&= \frac{1}{a^*} \Big( \lambda_1^2 + \frac{1}{\lambda_1^{p_1}} \inte
|x|^{p_1}Q^2(x)dx \Big)\\
&\geq \frac{2
}{a^*}\Big( \frac{p_1}{2}\inte |x
|^{p_1}Q^2(x)dx\Big)^\frac{2}{p_1+2}
,
\end{split}
\end{equation}
and the last equality holds if and only if
$$\lambda_1=\lam _0:=\Big( \frac{p_1}{2}\inte |x
|^{p_1}Q^2(x)dx\Big)^\frac{1}{p_1+2}.$$
On the other hand, it follows from $(3.20)$ in \cite{GS}  that
 \begin{equation}\label{lim}
\liminf_{(a_1,a_2)\nearrow (a^*,a^*)}\frac{e_1(a_1)}{\eps_1^{p_1}}=\frac{2
}{a^*}\Big( \frac{p_1}{2}\inte |x
|^{p_1}Q^2(x)dx\Big)^\frac{2}{p_1+2},
\end{equation}
Therefore, we can deduce from (\ref{mb}) and (\ref{lim}) that $\lambda_1\equiv\lam _0$, which gives (\ref{3:lm2}). Moreover, we also see that (\ref{str}) is indeed an equality, and thus $z_1=(0,0)$ follows, which implies (\ref{3:lm1}).

Furthermore, if  the limit $\mathbf{L}$ in (\ref{3:limit}) satisfies $ \mathbf{L}=0$, we then obtain from
(\ref{3:a_2}) that
\begin{equation}\label{3:ea1}
E^2_{a_2}(u_{a_2})=e_2(a_2)+o(1)e_2(a_2) \ \ \text{as}\,\
(a_1,a_2)\nearrow (a^*,a^*).
\end{equation}
Then, similar to the proof of (\ref{3:lm1}) and (\ref{3:lm2}), one can obtain the estimates (\ref{3:lm3}) and
(\ref{3:lm4}).

We finally remark that the above arguments can be proceeded for any subsequence of $\{(a_1,a_2)\}$ satisfying $(a_1,a_2)\nearrow (a^*,a^*)$,
 i.e., Theorem \ref{thm3.1} holds essentially for the whole sequence $\{(a_1,a_2)\}$, and the proof  of Theorem \ref{thm3.1} is therefore complete.\qed

In order to complete the proof of Theorem \ref{thm2}, in view of Theorem \ref{thm3.1} it next suffices to prove the following result for the case where the limit $\mathbf{L}$ in (\ref{3:limit}) is infinite.

\begin{thm}\label{thm3.3}
Suppose  that   (\ref{eq3.1}) holds, $\eps_1\geq\eps_2>0$, and  the limit $\mathbf{L}$ in (\ref{3:limit}) satisfies $ \mathbf{L}=\infty$.
 Then, all conclusions about $\{u_{a_1}\}$  in Theorem \ref{thm3.1} still hold, and  $\{u_{a_2}\}$ satisfies
\begin{equation}\label{3:a2limt2}
\lim_{(a_1,a_2)\nearrow (a^*,a^*)}  \tilde\eps_2 u_{a_2}(\tilde\eps_2
x+x_{a_2})=\frac{\lambda_2}{\|Q\|_{2}}Q(\lambda_2 x)
 \end{equation} strongly in $H^1(\R^2)$ for some $\lambda_2>0$, where
 \begin{equation}\label{3a:2limt2}
  \tilde\eps_2:=
\Big(\int_{\R^2}u_{a_2}^4dx\Big)^{-\frac{1}{2}}\to 0
\  \ \text{as}\,\ (a_1,a_2)\nearrow (a^*,a^*).
\end{equation}
Moreover, the  unique  maximum point ${x_{a_2}}$ of ${u_{a_2}}$ approaches $x_2$ as $(a_1,a_2)\nearrow (a^*,a^*)$.
\end{thm}

\noindent{\bf Proof.} Since the proof of this theorem is similar to
that of Theorem \ref{thm3.1}, we shall only sketch the main differences.
Because $ \mathbf{L}=\infty$, we first note from (\ref{3:a_2}) that
\begin{equation*}
\int_{\mathbb{R}^2}V_2(x)|u_{a_2}(x)|^2dx\leq
e_2(a_2)+C_2e^{\frac{-\delta_0}{\eps_1^{p_1}}} \leq Ce^{\frac{-\delta_0}{\eps_1^{p_1}}}\to 0 \ \ \text{as}\,\   (a_1,a_2)\nearrow (a^*,a^*).
\end{equation*}
Set
\begin{equation*}
\tilde\eps_2 ^{-2}(a_1,a_2):=\int_{\mathbb{R}^2}| u_{a_2}(x)|^4dx.
\end{equation*}
It then yields from (\ref{3:infinity}) that $\tilde\eps_2(a_1,a_2)\to 0$
as $(a_1,a_2)\nearrow (a^*,a^*)$. Moreover, since
\[\begin{split}
0&\leq \int_{\mathbb{R}^2}|\nabla
u_{a_2}(x)|^2dx-\frac{a_2}{2}\int_{\mathbb{R}^2}|
u_{a_2}(x)|^4dx= \int_{\mathbb{R}^2}| \nabla
u_{a_2}(x)|^2dx-\frac{a_2}{2}\tilde\eps_2^{-2}(a_2)\le e(a_1,a_2),
\end{split} \]
there exists a constant $m>0$, independent of $a_1$ and $a_2$, such that
\begin{equation*}
0<m\tilde\eps_2 ^{-2}(a_1,a_2)\leq\int_{\mathbb{R}^2}| \nabla
u_{a_2}|^2dx\leq \frac{1}{m}\tilde\eps_2 ^{-2}(a_1,a_2)\ \ \text{as
}\,\ (a_1,a_2)\nearrow (a^*,a^*).
\end{equation*}
On the other hand, the estimates of (\ref{eq2.8})-(\ref{eq2.10})
still hold for $u_{a_1}$. In view of above facts,
replacing
$\eps_2$ by $\tilde\eps_2$ and repeating   Steps 1-3 in the proof
of Theorem \ref{thm3.1}, one can obtain  (\ref{3:a2limt2}) and all  conclusions about $\{u_{a_1}\}$  in Theorem \ref{thm3.1} still holds,  which then completes the proof of Theorem \ref{thm3.3}.  \qed

\begin{rem}Since here $ \mathbf{L}=\infty$, we cannot get the estimate as
(\ref{3:ea1}) for $u_{a_2}$. Actually, we know only that $x_{a_2}\to
x_2$ as $a_2\nearrow a^*$, and  the explicit convergence
rate for $(\ref{3a:2limt2})$ remains   unknown in our Theorem \ref{thm3.3}.
\end{rem}

\section{Mass Concentration: Case of $x_1 =x_2$}

This section is devoted to the proof of Theorem \ref{thm3}. 
We always
denote $(u_{a_1},u_{a_2})$  a non-negative minimizer of
(\ref{eq1.3}), and our first result is concerned with the following energy estimates:

\begin{lem}\label{le4.1}
For $\beta >0$,  let  (\ref{eq1.18}) be satisfied and let $(a^*-a_1)\thicksim
(a^*-a_2)$ as $(a_1,a_2)\nearrow (a^*,a^*)$. Then there exists a positive  constant $C$, independent of $a_1$ and $a_2$, such that as $ (a_1,a_2)\nearrow (a^*,a^*)$,
\begin{equation}\label{4:est}
\begin{split}
\beta \int_{\R^2}u_{a_1}^2u_{a_2}^2dx+e_1(a_1)+e_2(a_2)&\leq
e(a_1,a_2)\\
&\leq
C(a^*-a_1)^\frac{p}{p+2}\Big(\ln\frac{1}{a^*-a_1}\Big)^\frac{2p}{p+2}
\end{split}\end{equation}
and for $i=1,\,2$,
 \begin{equation}
C(a^*-a_i)^\frac{p}{p+2}\le e_i(a_i) \leq  E_{a_i}^i(u_{a_i})\leq
C(a^*-a_i)^\frac{p}{p+2}\Big(\ln\frac{1}{a^*-a_i}\Big)^\frac{2p}{p+2}.
\label{4:ai}
\end{equation}
\end{lem}

\noindent{\bf Proof.} The lower bound of (\ref{4:est}) follows
directly from (\ref{3:left}). Inspired by the proof of Theorem \ref{thm1}, we next prove the upper bound of (\ref{4:est}) as follows: consider the trial function $(\phi_1,\phi_2)$ of the form (\ref{2:trial}) with $\bar x_1=\bar
x_2=x_0$, where $C_0>1$ is sufficiently large. By the exponential decay of $Q(x)$, we have
\begin{equation*}
\begin{split}
\int_{\R^2}V_i(x)\phi_i^2(x)dx&=\frac{1}{a^*}\big|\frac{\ln
\tau}{\tau}\big|^p\int_{\R^2}\Big|\frac{x}{\ln
\tau}-(-1)^iC_0\vec{n}\Big|^pQ^2(x)dx+O\big((R\tau)^{-\infty}\big)\\
&\leq 2C_0^p\big|\frac{\ln \tau}{\tau}\big|^p \ \ {\rm
as}\,\ R\tau \to \infty,\ \ i=1,\,2.
\end{split}
\end{equation*}
Together with (\ref{2:limt2}) and (\ref{2:limt5}), this implies that for $C_0>\frac{p+3}{2}$,
\begin{equation}\label{4:limt2}
\begin{split}
E(\phi_1,\phi_2)&\leq\sum_{i=1}^2\Big(\frac{a^*-a_i}{a^*}\tau^2+2C_0^p\big|\frac{\ln
\tau}{\tau}\big|^p\Big)+2\beta\tau^{2-2C_0}\\
&\leq
\sum_{i=1}^2\Big(\frac{a^*-a_i}{a^*}\tau^2+3C_0^p\big|\frac{\ln
\tau}{\tau}\big|^p\Big) \ \ {\rm as}\,\ \tau\to\infty.
\end{split}
\end{equation}
Since $a^*-a_1\sim a^*-a_2$ as $(a_1,a_2)\nearrow (a^*,a^*)$, setting $\tau=\big(\frac{1}{a^*-a_1}\big)^\frac{1}{p+2}\big(\ln\frac{1}{a^*-a_1}\big)^\frac{p}{p+2}$ into (\ref{4:limt2}) yields that
$$e(a_1,a_2)\leq E(\phi_1,\phi_2)\leq C(a^*-a_1)^\frac{p}{p+2}\Big(\ln\frac{1}{a^*-a_1}\Big)^\frac{2p}{p+2}$$
as $ (a_1,a_2)\nearrow (a^*,a^*)$. This gives the upper bound of (\ref{4:est}).

Finally, since
\[
E_{a_1}^1(u_{a_1})+E_{a_2}^2(u_{a_2})\le e(a_1,a_2)\text{ and } C(a^*-a_i)^\frac{p}{p+2}\le e_i(a_i) \leq  E_{a_i}^i(u_{a_i}), \ i=1,2,
\]
 we conclude (\ref{4:ai}) by applying (\ref{4:est}).
\qed

Based on Lemma \ref{le4.1}, a similar proof of Lemma \ref{le3.2}
gives the following  $L^4(\R^2)$ estimates of minimizers, and we
omit the details for simplicity.

\begin{lem}\label{le4.2}
Under the assumptions of  Lemma \ref{le4.1}, we have
\begin{equation}\label{4:L4}
\begin{split}
C(a^*-a_i)^{-\frac{2}{p+2}}\Big(\ln
\frac{1}{a^*-a_i}\Big)^{-\frac{4}{p+2}}&\leq
\int_{\R^2}u_{a_i}^4dx\\
&\leq
\frac{1}{C}(a^*-a_i)^{-\frac{2}{p+2}}\Big(\ln
\frac{1}{a^*-a_i}\Big)^{\frac{2p}{p+2}}
\end{split}\end{equation}
as $ (a_1,a_2)\nearrow (a^*,a^*)$, where $i=1,\,2$.
\end{lem}

Applying the above estimates, we are now ready to establish Theorem \ref{thm3} as follows.

\vskip 0.1truein

\noindent{\bf Proof of Theorem \ref{thm3}.} We establish this
theorem by three  steps:

\vskip 0.1truein

\noindent{\em Step 1: Scaling and limit behavior of minimizers $(u_{a_1}, u_{a_2})$.}
We first note from (\ref{4:ai}) that
\begin{equation}\label{4:v}
\int_{\mathbb{R}^2}V_i(x)|u_{a_i}(x)|^2dx\leq E^i_{a_i}(u_{a_i}) \le C(a^*-a_i)^\frac{p}{p+2}\Big(\ln\frac{1}{a^*-a_i}\Big)^\frac{2p}{p+2}
\end{equation}
as $(a_1,a_2)\nearrow (a^*,a^*)$, where $i=1,\,2$. Set
\begin{equation}\label{4:def}\epsilon ^{-2}_i(a_1,a_2):=\int_{\mathbb{R}^2}| u_{a_i}(x)|^4dx, \,\ i=1,\,2.
\end{equation}
It then yields from (\ref{4:L4}) that $\epsilon_i(a_1,a_2)\to 0$ as
$(a_1,a_2)\nearrow (a^*,a^*)$. Moreover, since
\begin{equation*}
\begin{split}
0&\leq\int_{\mathbb{R}^2}|\nabla
u_{a_i}(x)|^2dx-\frac{a_i}{2}\int_{\mathbb{R}^2}|
u_{a_i}(x)|^4dx\\
&=\int_{\mathbb{R}^2}| \nabla
u_{a_i}(x)|^2dx-\frac{a_i}{2}\epsilon^{-2}_i(a_1,a_2)\le e(a_1,a_2)
\ \ \text{as}\,\  (a_1,a_2)\nearrow (a^*,a^*),
\end{split}\end{equation*}
there exists a constant $m>0$, independent of $a_1$ and $a_2$, such that
\begin{equation}\label{4:def2}
0<m\epsilon_i ^{-2}(a_1,a_2) \leq \int_{\mathbb{R}^2}| \nabla
u_{a_i}|^2dx \leq \frac{1}{m}\epsilon _i^{-2}(a_1,a_2)\ \ \text{as}\,\
(a_1,a_2)\nearrow (a^*,a^*),
\end{equation}
where $i=1,\,2$.
We next denote $\epsilon _i(a_1,a_2)$ by $\epsilon_i$ for convenience.  Using
(\ref{4:v})-(\ref{4:def2}), similar to Step 1  in the  proof
of Theorem \ref{thm3.1}, one can prove that, if $x_{a_i}$ is a
global maximum point of $u_{a_i}$, then $x_{a_i}\to x_0$ as $(a_1,a_2)\nearrow (a^*,a^*)$. Define
\begin{equation}\label{4:scal}
\bar w_{a_i}=\epsilon_iu_{a_i}(\epsilon_ix+x_{a_i}),\,\ i=1,\,2.
\end{equation}
Then a proof similar to that of Theorem \ref{thm3.1} yields that $\bar w_{a_1}$ and $\bar w_{a_2}$ satisfy the elliptic system
\begin{equation*}
\arraycolsep=1.5pt
\begin{cases}
&-\Delta
\bar{w}_{a_1}+\epsilon_1^2 V_1(\epsilon_1
x+x_{a_1})\bar{w}_{a_1}
=\mu_{a_1}\epsilon_1^2 \bar{w}_{a_1}+a_1
\bar{w}_{a_1}^{3}-\beta(\frac{\epsilon_1}{\epsilon_2})^2\bar
w_{a_2}^2(\frac{\epsilon_1
x+x_{a_1}-x_{a_2}}{\eps_2})\bar w_{a_1}, \\
&-\Delta \bar{w}_{a_2}+\epsilon_2^2 V_2(\epsilon_2
x+x_{a_2})\bar{w}_{a_2}
=\mu_{a_2}\epsilon_2^2 \bar{w}_{a_2}+a_2
\bar{w}_{a_2}^{3}-\beta(\frac{\epsilon_2}{\epsilon_1})^2\bar
w_{a_1}^2(\frac{\epsilon_2 x+x_{a_2}-x_{a_1}}{\eps_1})\bar
w_{a_2},
\end{cases}\end{equation*}
where $(\mu_{a_1},\mu_{a_2})$ is a suitable Lagrange multiplier  and satisfies
\begin{equation}\label{4:scal2}
\lim_{\epsilon_i\to0}\bar
w_{a_i}(x)=\lim_{\epsilon_i\to0}\epsilon_iu_{a_i}(\epsilon_ix+x_{a_i})=
\frac{\lambda_i}{\|Q\|_2}Q(\lambda_ix)
\end{equation}
strongly in $H^1(\R^2)$ for some $\lam _i>0$, where $i=1,\,2$.
Moreover, by employing the comparison principle,   we have
\begin{equation}\label{44:exp} \bar w_{a_i}(x)\leq C
e^{-\frac{\lambda_i}{2}|x|} \ \ \text{for  large} \ \ |x|,
\end{equation}
and
\begin{equation}\label{44:lin}
\|\bar w_{a_i}(x)\|_{L^\infty(\R^2)}\leq C<\infty \ \
\text{as}\,\
(a_1,a_2)\nearrow (a^*,a^*),
\end{equation}
where $ C>0$ is independent of $a_1$ and $a_2$.

\vskip 0.1truein

\noindent{\em Step 2: Uniqueness of global maximum points $x_{a_i}$ for $u_{a_i}$.}
Stimulated by the proof of Theorem \ref{thm3.1}, the key of establishing the uniqueness of maximum points is to prove that for any  $ \Omega\subset\subset\R^2$, there exists a positive constant $C(\Omega)$ such that for any $\gamma\ge2$
\begin{equation}\label{4:comp1}
\Big(\frac{\epsilon_1}{\epsilon_2}\Big)^\gamma\bar w_{a_2}^2\big(\frac{\epsilon_1
x+x_{a_1}-x_{a_2}}{\epsilon_2}\big)\leq C(\Omega)
\quad
\text{uniformly\ in\ $\Omega$\ \ as}\,\ (a_1,a_2)\nearrow (a^*,a^*),\end{equation}
and
\begin{equation}\label{4:comp2}
\Big(\frac{\epsilon_2}{\epsilon_1}\Big)^\gamma\bar
w_{a_1}^2\big(\frac{\epsilon_2 x+x_{a_2}-x_{a_1}}{\epsilon_1}\big)\leq
C(\Omega) \quad
\text{uniformly\ in\ $\Omega$\ \ as}\,\  (a_1,a_2)\nearrow (a^*,a^*).\end{equation}
Indeed, if (\ref{4:comp1}) and (\ref{4:comp2}) hold,  by  the proof of Claim 2 of  Theorem \ref{thm3.1},  we know that $\bar{w}_{a_i}\rightarrow \bar{w}_i$ in $C^{2,\alpha}_{loc}(\mathbb{R}^2)$ for some $\alpha>0$, which  then gives the  uniqueness of global maximum points $x_{a_i}$ for $i=1,\,2$.

We next prove the estimates (\ref{4:comp1}) and (\ref{4:comp2}). We may assume that $\epsilon_1\geq\epsilon_2>0$ (passing to a subsequence if necessary). Then,
 (\ref{4:comp2}) directly follows from the $L^\infty$-boundedness
(\ref{44:lin}) of $\bar w_{a_1}$. As for (\ref{4:comp1}), it follows from (\ref{4:aa}) below, whose proof does not depend on  the uniqueness of $x_{a_2}$, that
for any $C>0$,
$$\frac{|\epsilon_1 x+x_{a_1}-x_{a_2}|}{\epsilon_2}> C\frac{\epsilon_1}{\epsilon_2} \ \
\text{uniformly\ in\ $\Omega$\ \ as}\,\ (a_1,a_2)\nearrow (a^*,a^*).$$
By the exponential decay (\ref{44:exp})  of $\bar
w_{a_2}$, we then derive that
$$\Big(\frac{\epsilon_1}{\epsilon_2}\Big)^\gamma\bar
w_{a_2}^2\big(\frac{\epsilon_1
x+x_{a_1}-x_{a_2}}{\epsilon_2}\big)\leq \Big(\frac{\epsilon_1}{\epsilon_2}\Big)^\gamma e^{-2C\frac{\epsilon_1}{\epsilon_2}}<C\ \
\text{uniformly\ in\ $\Omega$\ \ as}\,\  (a_1,a_2)\nearrow (a^*,a^*),$$
which therefore implies (\ref{4:comp1}).

\vskip 0.1truein

\noindent{\em Step 3: Proof of (\ref{thm3:3}).}
We first prove that \begin{equation}\label{4:aa}
\lim_{(a_1,a_2)\nearrow (a^*,a^*)}\frac{|x_{a_1}-x_{a_2}|}{\epsilon_i}=\infty,\ \
\mbox{where}\, \ i=1,\,2.
\end{equation}
Without loss of generality, as before we may assume
$\epsilon_1\geq\epsilon_2$ (passing to a subsequence if necessary), and it hence suffices to prove that
\begin{equation*}
\lim_{(a_1,a_2)\nearrow (a^*,a^*)}\frac{|x_{a_1}-x_{a_2}|}{\epsilon_1}=\infty.
\end{equation*}
On the contrary, suppose there exists a constant $R>0$ such that
$$\lim_{(a_1,a_2)\nearrow (a^*,a^*)}\frac{|x_{a_1}-x_{a_2}|}{\epsilon_1}\leq R<\infty.$$
We then have
\begin{equation}\label{4:R}
\Big|\frac{\epsilon_2}{\epsilon_1}x-\frac{x_{a_1}-x_{a_2}}{\epsilon_1}\Big|
\leq\frac{\epsilon_2}{\epsilon_1}R+R \leq 2R\ \ \mbox{for}\,\ x\in B_R(0)=\Big\{x\in\R^2:\, |x|\leq R\Big\}.\end{equation}
Since (\ref{4:scal2}) gives that
$$\bar w_{a_1}(x)\to\frac{\lambda_1}{\|Q\|_2}Q(\lambda_1x)\ \ \text{a.e.
 in} \,\ \R^2  \,\ \mbox{as}\,\ (a_1,a_2)\nearrow (a^*,a^*),$$
 we
derive from (\ref{4:R}) that
$$\lim_{(a_1,a_2)\nearrow(a^*,a^*)}\bar w_{a_1}\Big(\frac{\epsilon_2}{\epsilon_1}x-\frac{x_{a_1}-x_{a_2}}{\epsilon_1}\Big)\ge \inf_{x\in B_{2R}(0)}\frac{\lambda_1}{\|Q\|_2}Q(\lambda_1x) \geq \alpha>0\ \ \text{in} \,\ B_R(0).$$
By Fatou's Lemma, it then follows  from  (\ref{4:scal}) that
\begin{equation*}
\begin{split}
\int_{\R^2}u_{a_1}^2u_{a_2}^2dx&=\frac{1}{\epsilon_1^2\epsilon_2^2}\int_{\R^2}\bar w_{a_1}^2\Big(\frac{x-x_{a_1}}{\epsilon_1}\Big)\bar w_{a_2}^2\Big(\frac{x-x_{a_2}}{\epsilon_2}\Big)dx\\
&=\frac{1}{\epsilon_1^2}\int_{\R^2}\bar
w_{a_1}^2\Big(\frac{\epsilon_2}{\epsilon_1}x-\frac{x_{a_1}-x_{a_2}}{\epsilon_1}\Big)\bar w_{a_2}^2(x)dx\\
&\geq\frac{\alpha^2}{\epsilon_1^2}\int_{B_R(0)}\bar
w_{a_2}^2(x)>\frac{C}{\epsilon_1^2}\to \infty,
\end{split}
\end{equation*}
which implies that $\int_{\R^2}u_{a_1}^2u_{a_2}^2dx\to\infty$ as $(a_1,a_2)\nearrow (a^*,a^*)$, which leads to
a contradiction since the estimate (\ref{4:est}) implies that
$$\int_{\R^2}u_{a_1}^2u_{a_2}^2dx\leq
\frac{1}{\beta}e(a_1,a_2)\to 0 \ \ {\rm as}\,\ (a_1,a_2)\nearrow (a^*,a^*).$$
This completes the proof of (\ref{4:aa}).

We finally prove the estimate
\begin{equation}\label{4:aa2}
\limsup_{(a_1,a_2)\nearrow (a^*,a^*)} \frac{|x_{a_i}-x_0|}{\epsilon_i\ln
\frac{1}{\epsilon_i}}<\infty,\ \ \mbox{where}\,\ i=1,\,2.\end{equation} On the
contrary, without loss
of generality, we assume that (\ref{4:aa2}) is false for $i=1$. Then
 there exists a subsequence of $\{(a_1,a_2)\}$, still denoted by
$\{(a_1,a_2)\}$, such that
$$\lim_{(a_1,a_2)\nearrow (a^*,a^*)} \frac{|x_{a_1}-x_0|}{\epsilon_1\ln
\frac{1}{\epsilon_1}}=\infty.$$  By Fatou's Lemma, we deduce that for any $m>0$,
\begin{equation}\label{4:laa}
\begin{split}
\int_{\R^2}V_1(x)u_{a_1}^2(x)dx&=\int_{\R^2}|\epsilon_1x+x_{a_1}-x_0|^p\bar w_{a_1}^2(x)dx\\
&=\Big|\epsilon_1\ln
\frac{1}{\epsilon_1}\Big|^p\int_{\R^2}\Big|\frac{x}{\ln\frac{1}{\epsilon_1}}+\frac{x_{a_1}-x_0}{\epsilon_1\ln
\frac{1}{\epsilon_1}}\Big|^pw_{a_1}^2(x)dx\geq m\Big|\epsilon_1\ln
\frac{1}{\epsilon_1}\Big|^p,
\end{split}
\end{equation}
and hence
\begin{equation}\label{4:la}
\begin{split}
E^1_{a_1}(u_{a_1})&=\int_{\R^2}\big(|\nabla
u_{a_1}|^2+V_1(x)u_{a_1}^2\big)dx-\frac{a_1}{2}\int_{\R^2} u_{a_1}^4dx\\
&\geq \frac{a^*-a_1}{2}\int_{\R^2}
u_{a_1}^4dx+\int_{\R^2}V_1(x)u_{a_1}^2dx\\
&\geq \frac{a^*-a_1}{2}\frac{1}{\epsilon_1^2}+m\big|\epsilon_1\ln
\frac{1}{\epsilon_1}\big|^p \ \ \text{for any} \,\  m>0.
\end{split}
\end{equation}
Applying Lemma A in the appendix, we conclude from (\ref{4:la}) that
$$E^1_{a_1}(u_{a_1})\geq C(m,p)\big(a^*-a_1\big)^\frac{p}{p+2}\Big(\ln\frac{1}{a^*-a_1}\Big)^\frac{2p}{p+2},$$
where $ C(m,p)\to \infty $ as $m\to\infty.$ This however contradicts (\ref{4:ai}), and the proof of (\ref{4:aa2}) is therefore complete.

Since the arguments hold for any subsequence of $\{(a_1,a_2)\}$ satisfying $(a_1,a_2)\nearrow (a^*,a^*)$, Theorem \ref{thm3} holds  for the whole sequence $\{(a_1,a_2)\}$, and the proof  of Theorem \ref{thm3} is therefore complete.
\qed

\small

\section{Appendix: Some Proofs}

In this appendix, we give the proof of  Proposition A stated in the Introduction and establish a lemma which leads to the important estimate (\ref{thm3:3}) of Theorem \ref{thm3}.

\vskip 0.1truein

\noindent{\bf Proof of Proposition A.}
The energy estimate of (\ref{1:e}) follows directly from Lemma 3 of \cite{GS}. The uniqueness of maximum points of $\bar u_i(x)$ is proved in Theorem 1.1 of \cite{GZZ} for the case where $V_i(x)$ is ring-shaped, i.e.,
$V_i(x)=||x|-A|^2$, where $A>0$. One can however check that the argument in \cite{GZZ} is also applicable for the case where $V_i(x)$ takes the form of (\ref{1:V}).
It therefore remains to prove the estimate (\ref{1:dec}).

For convenience, we now consider  the case of  $i=1$, for which we set $w_1(x)=\eps_1\bar u(\eps_1 x+x_1)$. Note then from \cite{GS} that $w_1(x)$ satisfies
\begin{equation}\label{App1}
-\Delta w_1(x)+\eps_1^{2}V_1(\eps_1x+x_1)w_1(x)=\mu_1\eps_1^2w_1(x)+a_1w_1^3(x)\quad
\mbox{in}\,\ \R^2,
\end{equation}
where $\mu_1:=\mu_1(a_1)<0$ is a suitable Lagrange multiplier, and
$$\lambda_1^2:=-\lim_{a_1\nearrow a^*}\mu_1\eps_1^2=\Big(\frac{p_1}{2}\inte|x|^{p_1}Q^2(x)dx\Big)^\frac{2}{p_1+2}\,.$$
Moreover, it follows from \cite{GS} that $w_1(x)\to w_0$ strongly in $H^1(\R^2)$ as $a_1\nearrow a^*$, where $w_0>0$ is the unique radially symmetric solution of
$$-\Delta w_0(x)=-\lambda_1^2w_0(x)+a^*w_0^3(x)\quad
\mbox{in}\quad \R^2.$$
Hence, for any number $\alpha>2$,
\begin{equation}\label{App2}
\int_{|x|>R_0}|w_1|^\alpha dx\to 0\ \ \text{as}\,\   R_0\to \infty \quad \text{uniformly for}\,\  a_1\nearrow a^*.
\end{equation}
Note from (\ref{App1}) that $-\Delta w_1-c(x)w_1\leq0$ in $\R^2$, where $c(x)=a_1w_1^2$. By applying
De Giorgi-Nash-Moser theory, we  then  deduce from (\ref{App2})
that
\begin{equation*}
w_1(x)\rightarrow 0 \ \ \text{as}\,\   |x|\rightarrow\infty \quad
\text{uniformly for } a_1\nearrow a^*.
\end{equation*}
Since $w_1$ satisfies (\ref{App1}), one can use  the comparison
principle as in \cite{KW} to compare $w_1$ with
$e^{-\frac{\lambda_1}{2}|x|}$, which then shows that there exists a
large constant $R_1>0$, independent of $a_1$, such that
\begin{equation}\label{App3}
w_1(x)\leq e^{-\frac{\lambda_1}{2}|x|} \quad \text{for} \quad |x|>R_1
\ \ \text{as}\,\   a_1\nearrow a^*.
\end{equation}
For any $R>0$ and $x\in B^c_R(x_1)$, $\frac{|x-x_1|}{\eps_1}\geq\frac{R}{\eps_1}\to\infty$ as $a_1\nearrow a^*$. We thus obtain from (\ref{App3}) that for any $R>0$,
\begin{equation}\label{App4}
\bar u_1(x)=\frac{1}{\eps_1}w_1(\frac{x-x_1}{\eps_1})\leq \frac{1}{\eps_1}e^{-\frac{\lambda_1|x-x_1|}{2\eps_1}}\leq e^{-\frac{\lambda_1|x-x_1|}{4\eps_1}}\quad \text{in}\,\ B^c_R(x_1) \,\ \text{as}\,\ a_1\nearrow a^*.
\end{equation}
Similarly, the above argument also yields  that for any $R>0$,
 \begin{equation}\label{App5}
\bar u_2(x)\leq e^{-\frac{\lambda_2|x-x_2|}{4\eps_2}}\quad \text{in}\,\ B^c_R(x_2) \,\ \text{as}\,\ a_2\nearrow a^*,
\end{equation}
where $\lambda_2=\big(\frac{p_2}{2}\inte|x|^{p_2}Q^2(x)dx\big)^\frac{1}{p_2+2}$.
We therefore obtain the estimate (\ref{1:dec}) from (\ref{App4}) and (\ref{App5}) by setting $\delta=\min\{\lambda_1,\lambda_2\}>0$.
 \qed

The following lemma is used in the proof of Theorem \ref{thm3}.

\vskip 0.1truein

\noindent {\bf Lemma A.} {\em Consider positive constants $\kappa, m$ and $p$, and  $0<a<a^*$. Then for
the function
\begin{equation*}
f(s):=\frac{a^*-a}{\kappa}s^2+m\big|\frac{\ln s}{s}\big|^p, \quad where
\,\ s\in(e^3,\infty),
\end{equation*} there exists a positive constant $C:=C(\kappa,m,p)$, where $C(\kappa,m,p)\to \infty$ as $m\to\infty$, such that
\begin{eqnarray*}
f(s) \geq
C\big(a^*-a\big)^\frac{p}{p+2}\Big(\ln\frac{1}{a^*-a}\Big)^\frac{2p}{p+2} \quad{\rm
as}\,\ a\nearrow a^*\,.
\end{eqnarray*}}

\noindent{\bf Proof.} A direct calculation shows that if $a^*-a>0$ is sufficiently small, then
 $f(s)$ with $s\in(e^3,\infty)$ is strictly convex  and $f(s)$ has a unique global minimum point, denoted by $s_1$. Since $f'(s_1)=0$, we have
$$\frac{2(a^*-a)}{\kappa}s_1=mp\frac{(\ln
s_1)^p}{s_1^{p+1}}\Big(1-\frac{1}{\ln s_1}\Big)\,.$$ Note that $\frac{1}{\ln
s_1}<\frac{1}{3}$ for $s_1 > e^3$. Thus, one can derive from the
above equality that
$$\frac{2mp}{3}\frac{(\ln s_1)^p}{s_1^{p+1}}\leq\frac{2(a^*-a)}{\kappa}s_1\leq mp\frac{(\ln s_1)^p}{s_1^{p+1}},$$
which is equivalent to
\begin{equation}\label{4:func1}
\Big(\frac{mp\kappa}{3(a^*-a)}\Big)^\frac{1}{p+2}(\ln
s_1)^\frac{p}{p+2}\leq {s_1}\leq
\Big(\frac{mp\kappa}{2(a^*-a)}\Big)^\frac{1}{p+2}(\ln
s_1)^\frac{p}{p+2}.
\end{equation}
This implies that
$$\frac{a^*-a}{\kappa}s_1^2\geq\frac{a^*-a}{\kappa}\Big(\frac{mp\kappa }{3(a^*-a)}\Big)^\frac{2}{p+2}\big(\ln
s_1\big)^\frac{2p}{p+2}=\Big(\frac{a^*-a}{\kappa}\Big)^\frac{p}{p+2}\Big(\frac{mp}{3}\Big)^\frac{2}{p+2}\big(\ln
s_1\big)^\frac{2p}{p+2},$$
and
$$m\Big|\frac{\ln s_1}{s_1}\Big|^p\geq m\Big(\frac{a^*-a}{\kappa}\Big)^\frac{p}{p+2}\Big(\frac{2}{mp}\Big)^\frac{p}{p+2}\big(\ln
s_1\big)^\frac{2p}{p+2}\,.$$
Thus, there exists a positive constant $C(\kappa,p)$, independent of $a$, such that
\begin{equation}\label{4:func2}
f(s_1)=\frac{a^*-a}{\kappa}s_1^2+m\Big|\frac{\ln
s_1}{s_1}\Big|^p\geq
C(\kappa,p)m^\frac{2}{p+2}\big(a^*-a\big)^\frac{p}{p+2}(\ln
s_1)^\frac{2p}{p+2}\,.
\end{equation}
Note also from (\ref{4:func1}) that $$s_1\geq
\Big(\frac{mp\kappa}{3(a^*-a)}\Big)^\frac{1}{p+2} \text{ and }
 \ln s_1\geq
\frac{1}{p+2}\Big(\ln \frac{mp\kappa}{3}+\ln\frac{1}{a^*-a}\Big).$$
We therefore conclude from these estimates and (\ref{4:func2}) that for sufficiently large $m>0$,
\[\begin{split}
f(s_1)&\geq
C(\kappa,p)m^\frac{2}{p+2}\big(a^*-a\big)^\frac{p}{p+2}\Big(\ln
\frac{mp\kappa}{3}+\ln\frac{1}{a^*-a}\Big)^\frac{2p}{p+2} \\
&\geq \frac{C(\kappa,p)}{2}m^\frac{2}{p+2}\big(a^*-a\big)^
\frac{p}{p+2}\Big(\ln\frac{1}{a^*-a}\Big)^\frac{2p}{p+2},
\end{split}\]
which completes the proof of Lemma A.
\qed\\

\normalsize

\noindent {\bf Acknowledgements:}
The authors would like to thank the reviewer for her/his helpful comments upon which the paper was revised.   This research was  supported by NSFC under grants No. 11471331, 11501555 and 11671394.

\end{document}